\documentclass[arps,prd,onecolumn,showpacs]{revtex4}
\usepackage{graphicx}
\usepackage{amsmath}
\usepackage{amsfonts}
\usepackage{amssymb}
\usepackage{color}
\usepackage{epsf}

\newcommand{\be}{\begin{equation}}         
\newcommand{\ee}{\end{equation}}
\newcommand{\ba}{\begin{eqnarray}}
\newcommand{\ea}{\end{eqnarray}}
\newcommand{\nn}{\nonumber}
\newcommand{\rc}{\b {\it r}}
\newcommand{\tc}{\b {\it t}}

\newcommand\lsim{\mathrel{\rlap{\lower4pt\hbox{\hskip1pt$\sim$}}
        \raise1pt\hbox{$<$}}}
\newcommand\gsim{\mathrel{\rlap{\lower4pt\hbox{\hskip1pt$\sim$}}
        \raise1pt\hbox{$>$}}}

\def\x{{\bf x}}

\begin{document}

\title{Neutrino clustering around spherical dark matter halos}
\author{ Marilena LoVerde$^1$ and Matias Zaldarriaga$^2$}
\affiliation{$^{1}$ Enrico Fermi Institute, Kavli Institute for Cosmological Physics, Department of Astronomy and Astrophysics, University of Chicago, Illinois, 60637, U.S.A. \\
$^{2}$ School of Natural Sciences, Institute for Advanced Study, Princeton, New Jersey, 08540, U.S.A.\\ marilena@uchicago.edu, matiasz@ias.edu}
\begin{abstract}
Cold dark matter halos form within a smoothly distributed background of relic neutrinos -- at least some of which are massive and non-relativistic at late times. We calculate the accumulation of massive neutrinos around spherically collapsing cold dark matter halos in a cosmological background. We identify the physical extent of the ``neutrino halo" in the spherical collapse model, which is large in comparison with the virial radius of the dark matter,  and conditions under which neutrinos reaching the cold dark matter halo will remain bound to the halo at late times. We calculate the total neutrino mass and bound neutrino mass associated with isolated spherical halos for several neutrino mass hierarchies and provide fitting formulae for these quantities in terms of the cold dark matter halo mass and the masses of the individual neutrino species. 
\end{abstract}
\pacs{98.80.-k, 95.85.Ry,95.35.+d}

\maketitle

\section{Introduction}
\label{sec:intro}

Current cosmological data sets demonstrate that the large-scale properties of the Universe, the anisotropies in the cosmic microwave background (CMB) radiation for instance \cite{Ade:2013zuv,Hinshaw:2012aka,Hou:2012xq,Sievers:2013ica}, can be very accurately characterized by a flat, $\Lambda$ cold dark matter (CDM) cosmology.  The standard cosmological model predicts the existence of a cosmic neutrino background with the relic abundance of each of the three neutrino species comparable to the photon number density today ($\bar{n}_{1\nu} \approx 3/11\ \bar{n}_\gamma$) and with temperature $T_\nu \approx 1.95 K$ or $1.7 \times 10^{-4} eV$ (see, e.g. \cite{Lesgourgues:2006nd,Weinberg:2008zzc} for a review). The cosmic neutrino background contributes to the radiation density when the neutrinos are relativistic. Constraints on the relativistic degrees of freedom from CMB data and big bang nucleosynthesis are indeed consistent with three relativistic neutrino species at early times \cite{Ade:2013zuv,Hinshaw:2012aka,Hou:2012xq,Sievers:2013ica}, providing evidence for a neutrino contribution to the radiation density in the early Universe.  Neutrino oscillation data, however,  require that at least one of the neutrino species have mass $m_\nu \ge 0.048 eV$ and therefore at least one of the neutrino mass eigenstates behaves cosmologically as a nonrelativistic dark matter component today. The constraints on neutrino mass from Tritium $\beta$ decay in combination with the neutrino oscillation data require $m_{\nu i} \lsim 2 eV$ so that all three neutrino mass eigenstates are relativistic at least until the transition to matter domination \cite{Beringer:1900zz}. The influence of massive neutrinos on cosmological data is at present undetected. The upper bound on the sum of the neutrino masses from cosmology data ranges from $\sum_i m_{\nu i} \lsim 0.2 eV -1eV$, depending on the data set \cite{Reid:2009nq, Thomas:2009ae, Swanson:2010sk, Xia:2012na, RiemerSorensen:2011fe,Zhao:2012xw,dePutter:2012sh, Ade:2013zuv,Hinshaw:2012aka,Hou:2012xq,Sievers:2013ica}. Some analyses, however,  find that hints for massive neutrinos with $\sum_i  m_{\nu i}\sim 0.4eV$ may already be present in the data \cite{Wyman:2013lza}. 

\begin{figure}[t]
\begin{center}
$\begin{array}{cc}
\includegraphics[width=0.5\textwidth]{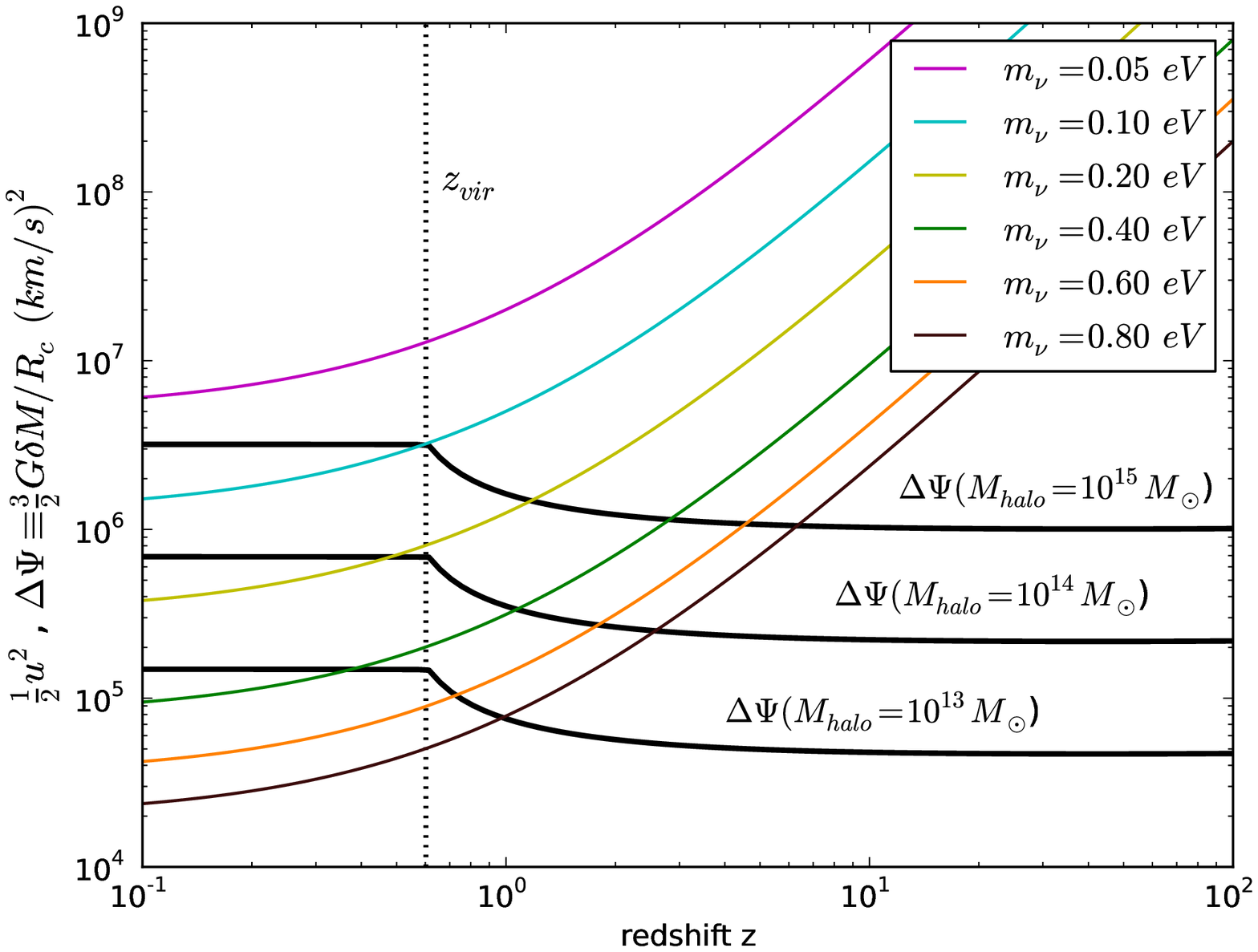}  & \includegraphics[width=0.5\textwidth]{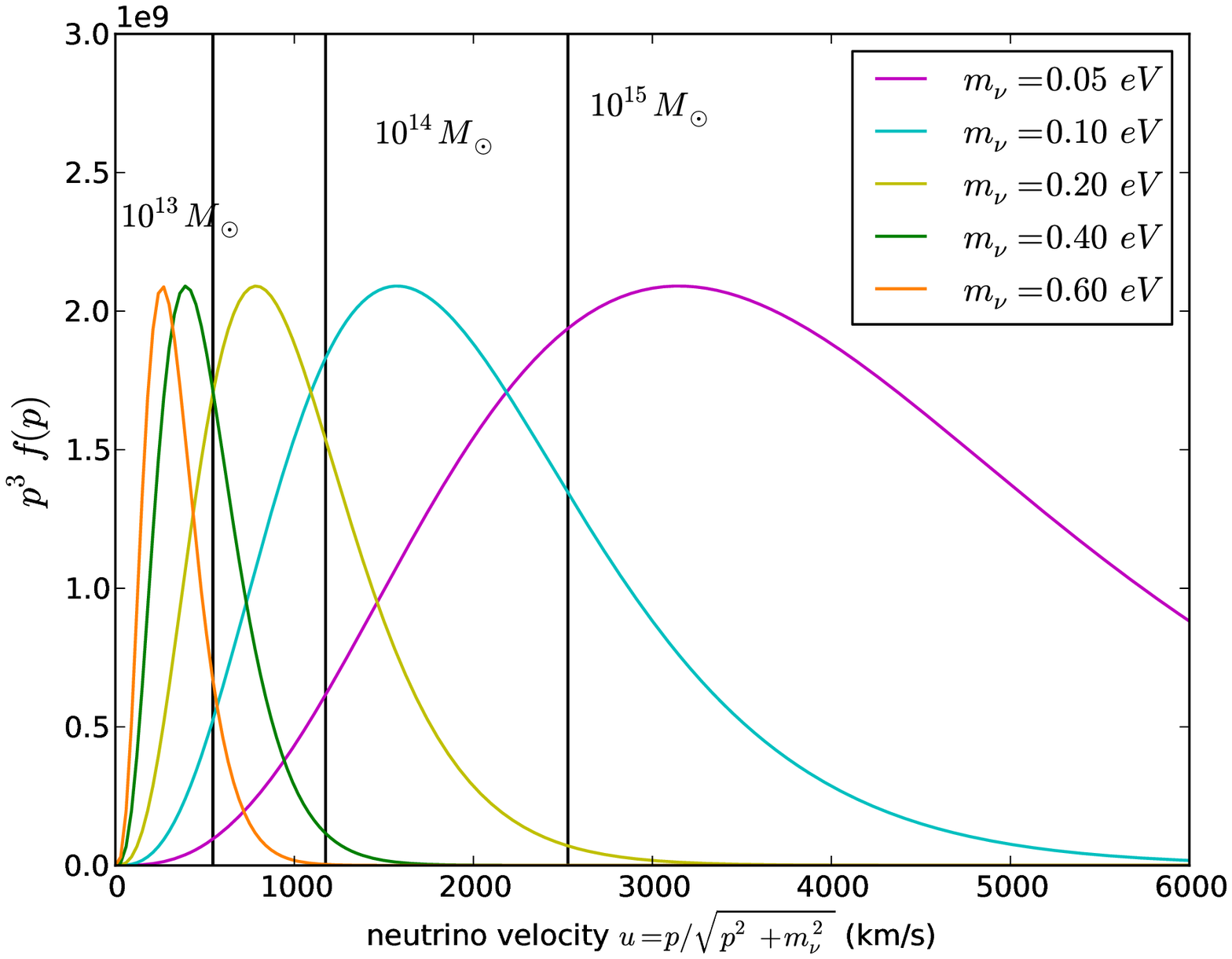}  
\end{array}$
\caption{\label{fig:Phivpec}Comparison of  the peculiar velocity of neutrinos to the depth of the gravitational potential well for different halo masses.  Left:  Thick solid curves are $\Delta\Psi \equiv \frac{3}{2} G\delta M/R_c$ for spherical-top hat halos with $M = 10^{13}M_\odot$, $ = 10^{14}M_\odot$, $= 10^{15}M_\odot$ that virialize at roughly the same time (indicated by the vertical line). For each neutrino mass we plot $\frac{1}{2}u^2$ where $u = 3.151 T_\nu/m_\nu$ -- the average magnitude of the peculiar velocity for neutrinos drawn from a Fermi-Dirac distribution with temperature $T_\nu = 1.95 K$.  Right: The distribution of neutrino velocities today for different neutrino masses (in the non-relativistic limit $p = m_\nu u$). Black vertical lines show the depth the potential wells $\sqrt{2\Delta\Psi}$.  }  
\end{center} 
\end{figure}

Any primordial fluctuations in the neutrino density will be washed out on scales below the neutrino free-streaming scale \cite{Bond:1980ha,Davis:1981yf}. Nevertheless, even for the minimum mass (of the most massive) neutrino, $m_\nu \sim 0.05 eV$,  a significant fraction of neutrinos will have velocities comparable to, and less than, the escape velocity for massive halos today allowing them to participate in gravitational clustering at late times (Fig.~\ref{fig:Phivpec}). Moreover, we shall see that even at earlier times some fraction of the neutrino population is moving slowly enough to be captured by a collapsing halo. 
 
 In this paper we explore the accretion and clustering of massive neutrinos around CDM halos. We restrict to the simplest possible model of a cold dark matter halo: a spherical top-hat density profile following the usual spherical collapse solution solved in a $\nu \Lambda CDM$ background \cite{Gunn:1972sv} (see \cite{Ichiki:2011ue} for inclusion of massive neutrinos). The spherical top-hat density profile is of course an unrealistic description of a dark matter halo, but is nevertheless useful because it provides a complete description of dark matter halo formation in a cosmological context. Within the spherical collapse model, we determine the boundary of the ``neutrino halo"-- which extends to a much greater radius than the virial radius of the CDM, calculate the neutrino mass clustered around the CDM halo, the neutrino mass interior to the virial radius, and the neutrino mass that remains bound within the halo into the $\Lambda$-dominated era.  Our results can be used to estimate the neutrino mass around CDM halos (we provide fitting formulae in Eq.~(\ref{eq:Mnurstarfit}) and Eq.~(\ref{eq:Mnuboundfit})) and incorporate some effects of non-linear clustering of massive neutrinos into standard spherical collapse calculations. 

Before proceeding we discuss some related literature. Using the approach of Gilbert \cite{1966ApJ...144..233G}, Brandenberger, Kaiser, and Turok  (BKT) \cite{Brandenberger:1987kf} used a linearized solution to the Boltzmann equation to determine the density profile of neutrinos around cosmic strings. The BKT method was applied by Singh and Ma \cite{Singh:2002de} to study the neutrino density profile around cold dark matter halos described by a Navarro, Frenk, and White (NFW) profile \cite{Navarro:1995iw}. This method is one of several approaches we take to computing the neutrino mass around our halos in \S\ref{sec:clustering}.  Abazajian et al \cite{Abazajian:2004zh} applied the BKT methodology to determine the neutrino density profile for their halo model of the CDM plus neutrino power spectrum. While we focus on individual halos, it is worth noting analytic studies of the halo and matter power spectra in mixed dark matter ($\nu+ CDM$) cosmologies: Saito et al \cite{Saito:2009ah}, as well as Wong \cite{Wong:2008ws} and Upadhye et al \cite{Upadhye:2013ndm}  developed perturbation theory techniques for this purpose. More closely related is Ringwald and Wong \cite{Ringwald:2004np}. Ringwald and Wong performed a comprehensive calculation of massive neutrino clustering around static NFW profiles with a more accurate calculation than the BKT method, dubbed the ``N-1-body" method, and we take a similar approach in \S \ref{ssec:full}. Ringwald and Wong noted that the linearized approximation of \cite{Brandenberger:1987kf,Singh:2002de} underestimates the clustering of neutrinos for very massive halos and/or neutrinos with large masses and our results are in agreement. 

A number of authors have developed different techniques for studying the effects of massive neutrinos on CDM structure growth in N-body simulations: (i) through modification of the initial CDM power spectrum and background evolution (see e.g. \cite{Agarwal:2010mt, Upadhye:2013ndm} and references therein) (ii) using hybrid perturbation theory +  N-body schemes (e.g. \cite{Brandbyge:2008rv,Viel:2010bn,Marulli:2011he,AliHaimoud:2012vj}), (iii) and/or by starting the simulations at late times and directly including warm dark matter particles sampled from the neutrino phase space (e.g. \cite{Bode:2000gq,Colin:2007bk,Brandbyge:2008rv,Viel:2010bn,Brandbyge:2010ge,VillaescusaNavarro:2012ag}). Where applicable, we make comparisons with the recent work of Villaescusa-Navarro et al, \cite{VillaescusaNavarro:2012ag}, who use CDM + neutrino particles in N-body simulations to study clustering of massive neutrinos around CDM halos at late times. 

N-body simulations are the community standard for modeling and interpreting data. Nevertheless, semi-analytic descriptions of structure formation, such as spherical collapse, remain useful because they aid understanding of different physical effects and can more readily be extended to alternative cosmological models.  Our main goal here is to model the behavior of massive neutrinos around an individual halo through the process of halo formation, albeit within the simplified spherical collapse model.  Our results are therefore complementary to  \cite{Singh:2002de,Ringwald:2004np}, who focused on determining the local density of neutrinos within static halos and to \cite{VillaescusaNavarro:2012ag} who exclusively use N-body methods.

An outline of the paper is as follows.  In \S \ref{sec:dynamics}, we study the dynamics of individual neutrinos near a spherical top-hat halo and determine the physical extent of the neutrino halo. In \S \ref{sec:clustering} we  determine the neutrino mass and bound neutrino mass interior to the boundary of the neutrino halo and the boundary  of the CDM density perturbation using several different methods: an approximate solution of the Boltzmann equation \S\ref{ssec:BKT}, an ``absorbing barrier" model for the accretion of bound neutrino mass \S \ref{ssec:accrete}, and finally an exact Boltzmann calculation for neutrinos in an external halo in \S \ref{ssec:full}.  In \S \ref{sec:totalmass} we present our final results for the total mass of the  neutrino halo as a function of CDM halo mass for different neutrino mass hierarchies. We conclude in \S \ref{sec:conclusion}. Throughout the body of this paper we make the approximation that neutrinos that cluster can be treated by Newtonian mechanics: relativistic expressions and justification of the Newtonian limit is given in Appendix \S \ref{sec:GRNewton}. A discussion of what, precisely, we take as the definition of  a neutrino being ``bound" to a halo is in Appendix \S \ref{sec:bound}.

In the plots and numerical examples shown throughout this paper with use Planck \cite{Ade:2013zuv} values of the standard, flat $\Lambda CDM$ cosmological parameters: Hubble parameter $h = 0.67$, cold dark matter (CDM) density  $\Omega_c h^2 = 0.1199 $ and baryon density $\Omega_bh^2 =  0.022 $.  In the calculations throughout this paper we treat baryons as $CDM$, that is $\left. \Omega_c\right|_{{\rm {\tiny this\, paper}}} = \left. (\Omega_c + \Omega_b)\right|_{\rm{\tiny true}}$.We assume three species of massive neutrinos with variable masses $m_{\nu 1}$, $m_{\nu 2}$ and $m_{\nu 3}$. Massive neutrinos contribute a fraction $\Omega_{\nu }h^2 \approx \sum_i m_{\nu i}/(94 eV)$ to the critical energy density so for fixed CDM and baryon densities, changing the neutrino masses leads to a different total matter ($\Omega_m = \Omega_c + \Omega_b +  \Omega_\nu$) density today. We adjust $\Omega_\Lambda$ to keep the Universe flat, that is $\Omega_\Lambda = 1 - \Omega_c - \Omega_b - \Omega_\nu - \Omega_\gamma$.  We consider several representative scenarios for the neutrino masses that are approximately compatible with neutrino oscillation data. We solve for the background cosmology and spherical halo collapse self-consistently for each set of neutrino masses. The sets of neutrino masses we consider are: ``normal hierarchical" $m_{\nu 1} = 0.05eV$, $m_{\nu 2} = 0.01 eV$, $m_{\nu 3} = 0.0 eV$; ``inverted hierarchical" $m_{\nu 1} = 0.00eV$, $m_{\nu 2} = 0.05 eV$, $m_{\nu 3} = 0.05 eV$; and number of ``degenerate" scenarios: $m_{\nu 1} = m_{\nu 2} = m_{\nu 3} = 0.10,\,  0.20,\,  0.40,\, 0.60,\, 0.80 eV$  \cite{Beringer:1900zz}.   

For the halo evolution we use the spherical collapse solution for $R_c(t)$, a sphere enclosing a constant mass $M$ of cold dark matter, calculated for a general $\nu \Lambda CDM$ cosmology with the parameters described above. Our method of calculating $R_c(t)$ largely follows \cite{Naoz:2006ye, Ichiki:2011ue}: on super-horizon scales we use the perturbed Friedmann equation with adiabatic initial conditions for the density perturbations in each component; we use this solution up until horizon crossing, and from then on we use the acceleration equation for $R_c(t)$. During the subhorizon evolution we neglect perturbations in anything other than CDM (for exact details see \cite{LoVerde:2014rxa}). In this paper we are interested in neutrino accretion during and after halo collapse, but the spherical collapse solution sends $R_c(t_{collapse}) \rightarrow 0$, which is unphysical.  We handle this as follows: for $t\ge t_{vir}$ we set $R_c(t) = const. = R_{vir}$, where $R_{vir} = R_{max}/2$ and $R_c(t_{vir})\equiv R_{vir}$. We choose this form for the halo (constant $R_{c}$ after virialization) so that the CDM mass density inside the halo is constant after virialization. We note that the neutrinos reaching the halo will have originated at distances $r\gg R_c$, so for most of their journey they are completely insensitive to precisely how we treat the halo potential interior to $R_c$; i.e. their equation of motion just sees $\nabla \Psi \sim G M/r^2{\hat{r}}$. 

\section{Dynamics of a single neutrino}
\label{sec:dynamics}

For a neutrino trajectory to be significantly perturbed by a dark matter halo it must be non-relativistic so we can safely use the Newtonian equation of motion (see Appendix \S \ref{sec:GRNewton} for a more in-depth discussion of the relativistic and Newtonian equations of motion). The equation of motion for a non-relativistic particle in an expanding Universe is, 
\be
\label{eq:NREOM}
\frac{d{\bf v}}{dt}=-\nabla_{r} \left(\Psi_H(r,t)+\Psi_{pec}(r,t)\right)
\ee
where $r$ is the proper distance, $v$ is the physical velocity (including both Hubble and peculiar velocity which we denote by $u$), $\Psi_H$ is the potential due to the Hubble flow, $\Psi_{pec}$ is the peculiar gravitational potential, and we've neglected terms proportional to $\dot{\Psi}$. The Hubble potential is given by 
\be
\label{eq:PsiHubble}
\Psi_H(r,t) =-\frac{1}{2}\left(\dot{H}+H^2\right)r^2
\ee
where $\dot{}$ indicates the derivative with respect to time. The Hubble potential vanishes only for a particle at $r=0$.  For a spherical top-hat CDM density perturbation of mass $M$ and (proper) radius $R_c$ the peculiar gravitational potential is 
\be
\label{eq:Psipec}
\Psi_{pec}(r,t)=\left\{\begin{array}{ll}
\frac{G\delta M}{2}\frac{r^2}{R_c^3}& \textrm{ for } r \le R_c\\
-\frac{G\delta M}{r} +\frac{3}{2}\frac{G\delta M}{R_c}& \textrm{ for } r > R_c
\end{array}\right. 
\ee
where $\delta M = M-\frac{4}{3}\pi R_c^3 \rho_c$ and $\rho_c$ is the mean density of cold dark matter. Note that we have made the non-standard choice of setting the potential to zero at $r=0$, as opposed to $r =\infty$. We have done this so that net (Hubble plus peculiar) potential  is defined as the amount of work done in moving a particle from the origin to a distance $r$ at a fixed time. Interior to $R_c$, the total potential due to CDM is just $\left.\Psi_H+ \Psi_{pec}\right|_{CDM\, only} =   \frac{1}{4}H^2r^2\Omega_c(1+\delta)$ where $\delta \equiv \delta M/(\frac{4}{3}\pi R_c^3\rho_c)$ and $\Omega_c$ is the critical density of CDM. Alternatively, we can write $\left.\Psi\right|_{CDM\, only} = \frac{3}{2}GMr^2/R_c^3$ which makes it clear that inside of a virialized halo ($R_c = const$) the potential is constant if only CDM is present.

\begin{figure}[t]
\begin{center}$\begin{array}{cc}
\includegraphics[width=0.5\textwidth]{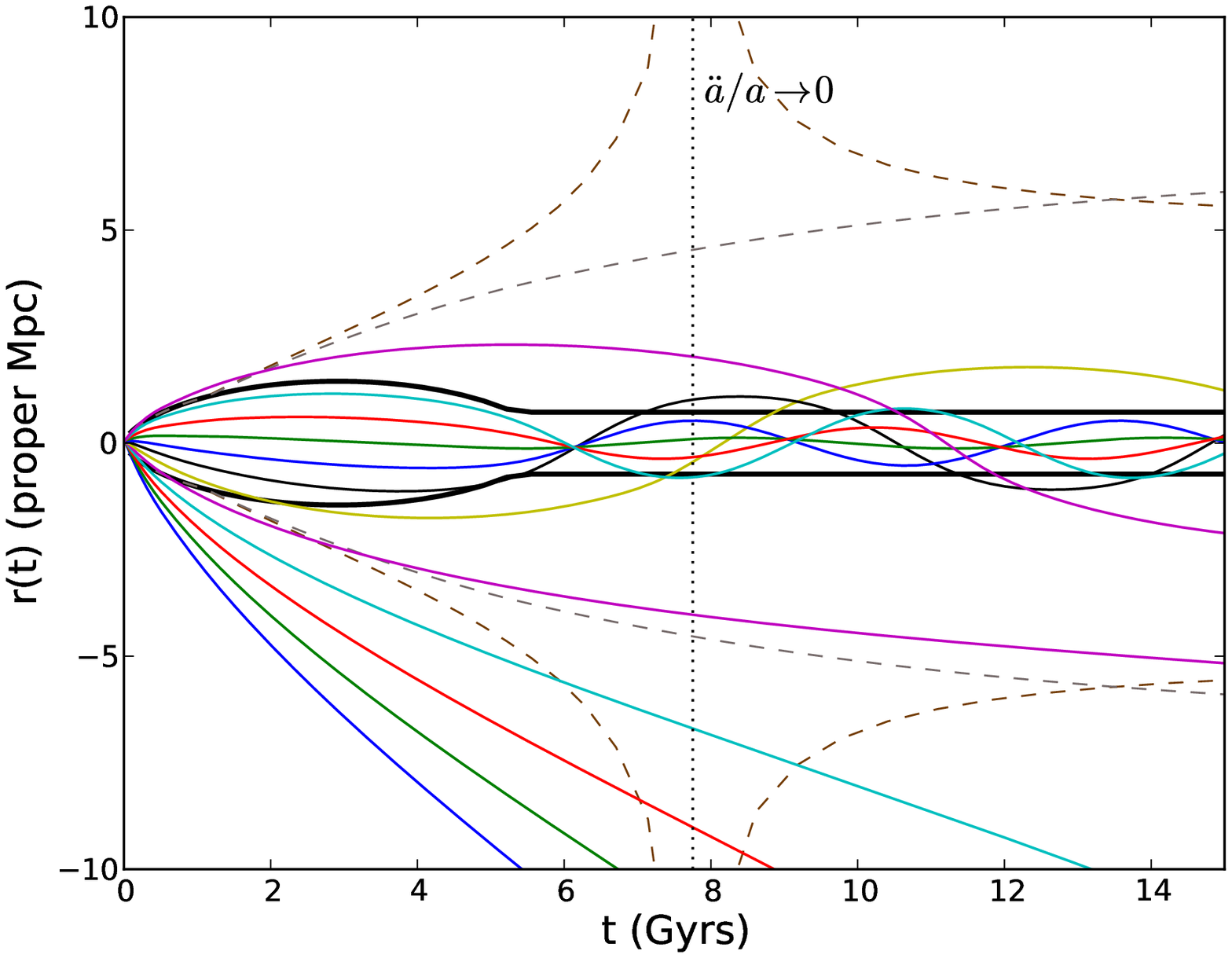} &\includegraphics[width=0.5\textwidth]{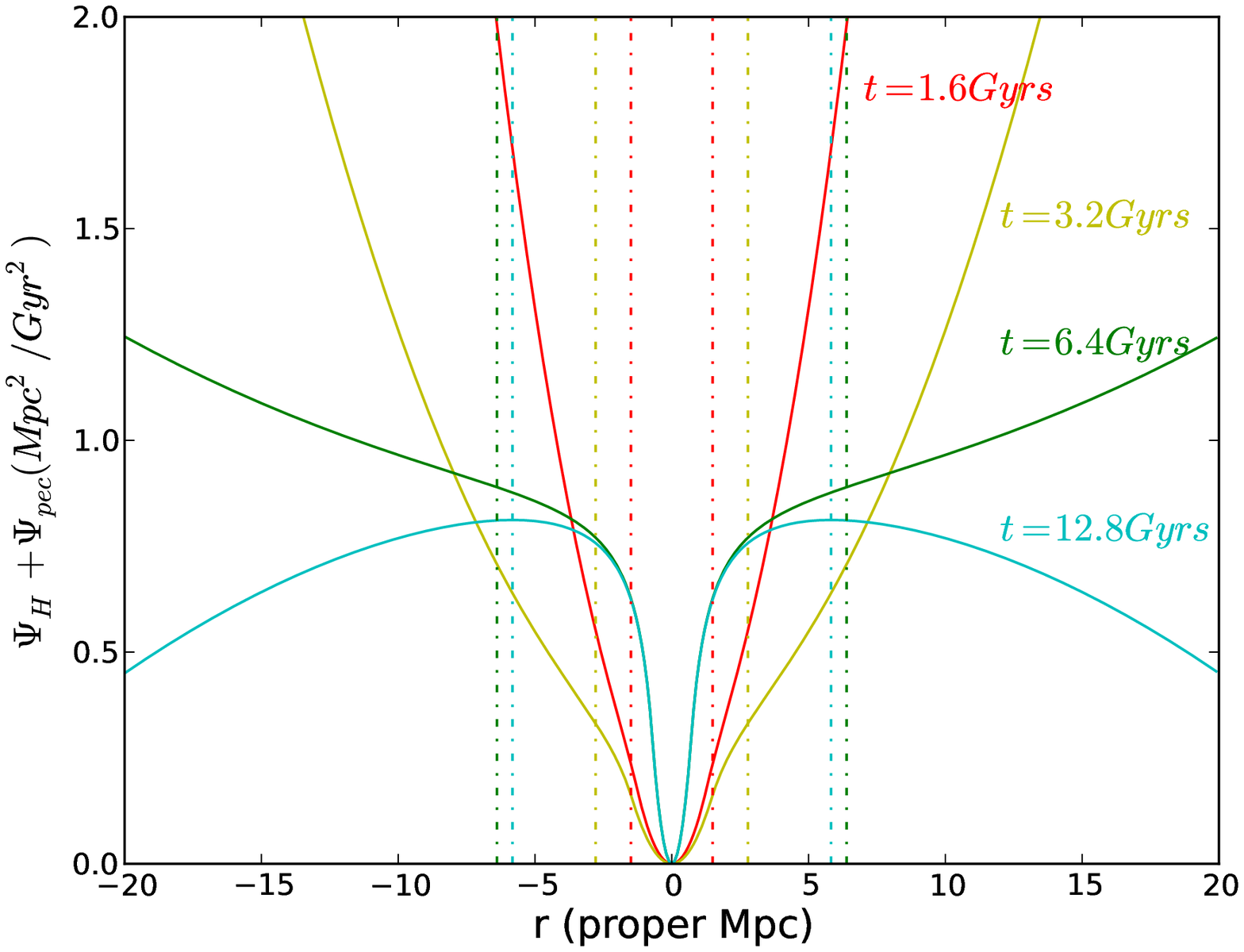} 
\end{array}$
\caption{\label{fig:rofts} Left: Solid colored lines are trajectories of $m_\nu= 0.05 eV$ neutrinos on radial paths through a collapsing halo of mass $M=10^{14} M_\odot$. The different colors indicate neutrinos with different velocities at the time that they reach $R_\nu$. The radius of the collapsing halo is shown in the thick black line (after $t\sim 6$ Gyrs the halo is assumed to have constant proper size $R_{virial}$). The time at which $\dot{H}+H^2\rightarrow 0$ is indicated by the vertical dotted line. The dashed brown line (that diverges when $\ddot{a}/a\rightarrow 0$) is $R_\nu(t)$, the scale at which the gravitational force of the halo dominates over the background (see Eq.~(\ref{eq:Rnu})) and another scale $r_*(t)$ is shown in the dashed gray line.  Right: The total gravitational potential (the sum of  $\Psi_H$ in Eq.~(\ref{eq:PsiHubble}) and $\Psi_{pec}$ in Eq.~(\ref{eq:Psipec})) plotted at several different times.  $R_\nu(t)$ at each time is indicated by the dashed vertical line of the same color.}  
\end{center}
\end{figure}


Substituting Eq.~(\ref{eq:PsiHubble}) and Eq.~(\ref{eq:Psipec}) into Eq.~(\ref{eq:NREOM})  we see that when a particle is within $R_\nu$ of the center of the halo, the force of the potential due to the halo dominates over the Hubble expansion where 
\be
\label{eq:Rnu}
R_\nu=\left|\frac{\Omega_c\delta}{2\left(1+\dot{H}/H^2\right)}\right|^{1/3}R_c 
\ee
which has a particularly simple form in Einstein de Sitter ($\Omega_m=1$), 
\be
\left.R_\nu\right|_{EdS} = \delta^{1/3} R_c
\ee
For a virialized halo during matter domination $\delta\sim200$ and $1+\dot{H}/H^2=-0.5$ giving $R_\nu\sim 8 R_c$. For our assumed potential in Eq.~(\ref{eq:Psipec}), a matter-dominated solution for $R_\nu$ (a radius at which the halo potential dominates over the background during matter domination \footnote{By ``matter domination" we actually mean a slightly stricter condition than $\Omega_m > \Omega_\Lambda$,  but the condition $\Omega_m > 2\Omega_\Lambda$ so that $\ddot{a}/a <0$.}) exists only for $R_\nu> R_c(t)$. Note that the presence of the cosmological constant qualitatively changes the behavior of particles approaching the halo: at the transition between matter domination and $\Lambda$ domination, $1+\dot{H}/H^2$ passes through zero sending $R_\nu \rightarrow \infty$. At late times in a $\Lambda CDM$ Universe, the force of $\nabla \Psi_H$ opposes $\nabla \Psi_{pec}$ and $R_\nu$ indicates the radius at which the two forces cancel. At late times (after $\ddot{a}/a >0$) $R_\nu$ is an upper bound for orbital radii of particles bound to the halo.

In Fig.~\ref{fig:rofts} we plot the solutions to Eq.~(\ref{eq:NREOM}) for particles on radial trajectories through the center of a collapsing halo. We assume a spherical top-hat density profile for the dark matter halo defined by radius $R_c(t)$ that follows the usual spherical collapse solution (as described in \S \ref{sec:intro}). Also plotted are the scale at which the gravitational force of the halo dominates over the background, $R_\nu(t)$, and another scale $r_*$, defined through 
\be
\label{eq:rstar}
r_*= \left\{\begin{array}{ll} \left|\Omega_c\delta \right|^{1/3}R_c & {\rm for} \quad \Omega_c\delta \ge 1\\
R_c(t) & {\rm else}\end{array}\right.\,.
\ee
For $\Omega_c\delta\ge 1$ this definition is equivalent to $\Psi_{pec}(r_*,t)-\Psi_{pec}(r=\infty,t)=\frac{1}{2}r_*^2H^2$, that is the radius at which the peculiar velocity due to the halo is equal to the Hubble flow. Note that for $\Omega_c\delta\ge1$, 
\be
\frac{dr_*}{dt}=\left(1+\frac{P}{\rho}+\frac{1}{\delta}\left(1-H^{-1}\frac{d\ln R_c}{dt}\right)\right)Hr_*
\ee
and 
\be
\frac{dR_\nu}{dt}=\left(\frac{1}{\delta}\left(1-H^{-1}\frac{d\ln R_c}{dt}\right)-\frac{H^{-1}}{3}\frac{d\ln \frac{\ddot{a}}{a}}{dt}\right)HR_\nu
\ee
So, for $\delta \gg1$,  $r_*$ and $R_\nu$ grow with the Hubble flow during matter domination and approach constants during $\Lambda$ domination. 

Throughout this paper we use $r_*$ to characterize the scale of the neutrino halo.  We note that while we sometimes refer to the scale $r_*$ as the ``boundary of the neutrino halo,"  this is a slight abuse of notation because $r_*$ does not actually depend on any neutrino properties -- it is determined solely by $\Psi_{pec}$ and $\Psi_H$. Another important point is that $r_*$ describes the radius of the sphere of gravitational influence of an isolated CDM halo in an expanding universe, but real halos are not isolated objects. In a cosmological context, the gravitational effects of nearby structure may well become important within $r_*$ and truncate the neutrino halo at a smaller radius.

\section{Neutrino clustering around a spherical halo}
\label{sec:clustering}
In this section we study the clustering of neutrinos around dark matter halos. Throughout we assume that gravitational effects of neutrino clustering can be neglected when calculating the neutrino trajectories. We shall see that even in the most clustered cases the neutrino mass remains below a few percent of the CDM mass interior to the halo ($<R_c$)  so this approximation should be justified. In this limit, the clustering of each neutrino mass eigenstate can be considered separately. In the example plots in this section we calculate the background cosmology (and evolution of the CDM halo radius $R_c$) for a cosmology that includes massive neutrinos with the following mass spectra: for $m_{\nu i } \ge 0.1 eV$, we have assumed a degenerate spectrum with $m_{\nu 1} = m_{\nu 2} = m_{\nu 3}$, and the plots showing $m_{\nu i } = 0.05$ assume the masses follow a normal hierarchy $m_{\nu 1} = 0.05 eV$, $m_{\nu 2} = 0.01 eV$ and $m_{\nu 3} = 0eV$.

For a single species of non-relativistic neutrino with mass $m_\nu$, the total neutrino mass interior to a proper radius $r$ is given by 
\ba
\label{eq:Mnultrstar}
M_\nu(<r)= m_\nu  \int_{V_x}d ^3{\bf x}\, a^3(t)\int_{V_p} \frac{d^3{\bf p}}{(2\pi)^3}  \, f({\bf x},{\bf p},t)
\ea
where $\x$ is comoving position, ${\bf p}$ the momentum ($p =  m_\nu {\bf u}/\sqrt{1-u^2} \approx m_\nu {\bf u}$ where ${\bf u} = {\bf v}- H {\bf r}$ is the peculiar velocity), $V_x=\frac{4}{3}\pi r^3(t)/a^3(t)$ is the comoving volume, and $V_p$ is the (infinite) volume in momentum space. At late times the neutrino distribution function, $f({\bf x}, {\bf p}, t)$, satisfies the non-relativistic Boltzmann equation: 
\be
\label{eq:Boltzmann}
\frac{\partial f}{\partial t} + \frac{{\bf p}}{a m_\nu}\cdot \nabla_x f - \left({\bf p} (H - \dot\Psi_{pec}) + m_\nu/a\nabla_x \Psi_{pec}\right)\cdot \nabla_p f = 0\,.
\ee
Hereafter we neglect the $\dot\Psi_{pec}$ term (see Appendix \ref{sec:GRNewton}). 

The unperturbed neutrino distribution function is given by the relativistic Fermi-Dirac distribution leftover from decoupling:
\be
\label{eq:f0}
f_0(p,a) \equiv \frac{2}{e^{a p/T_\nu} +1}\qquad {\rm so}\quad \bar{n}_{1\nu} = 3\zeta(3) T_\nu^3/(2\pi^2) \approx 112/cm^3
\ee
where $T_\nu \approx 1.95 K$ is the neutrino temperature today (with $a=1$) and $\bar n_{1\nu}$ is the comoving number density of one neutrino and anti-neutrino species. In the next few subsections we use different methods to recover expressions for the perturbed neutrino distribution function, $f({\bf x}, {\bf p}, t) = f_0(p, a) + f_1({\bf p}, {\bf x}, a)$ so that we may determine $M_\nu (<r)$.

\subsection{Approximate solution to the Boltzmann equation from BKT}
\label{ssec:BKT}
 \begin{figure}
\begin{center}$\begin{tabular}{cc}
\includegraphics[width=0.5\textwidth]{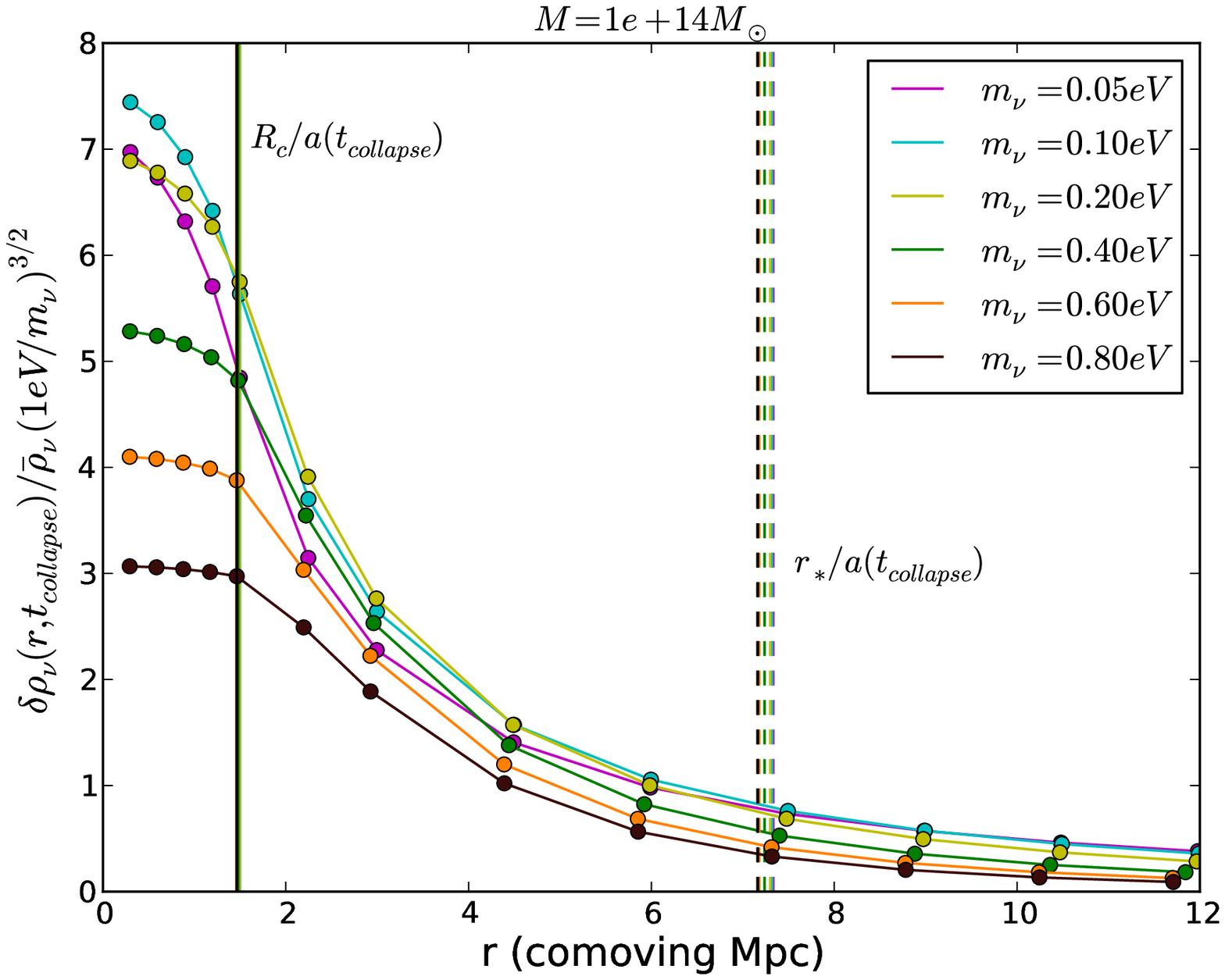} &\includegraphics[width=0.5\textwidth]{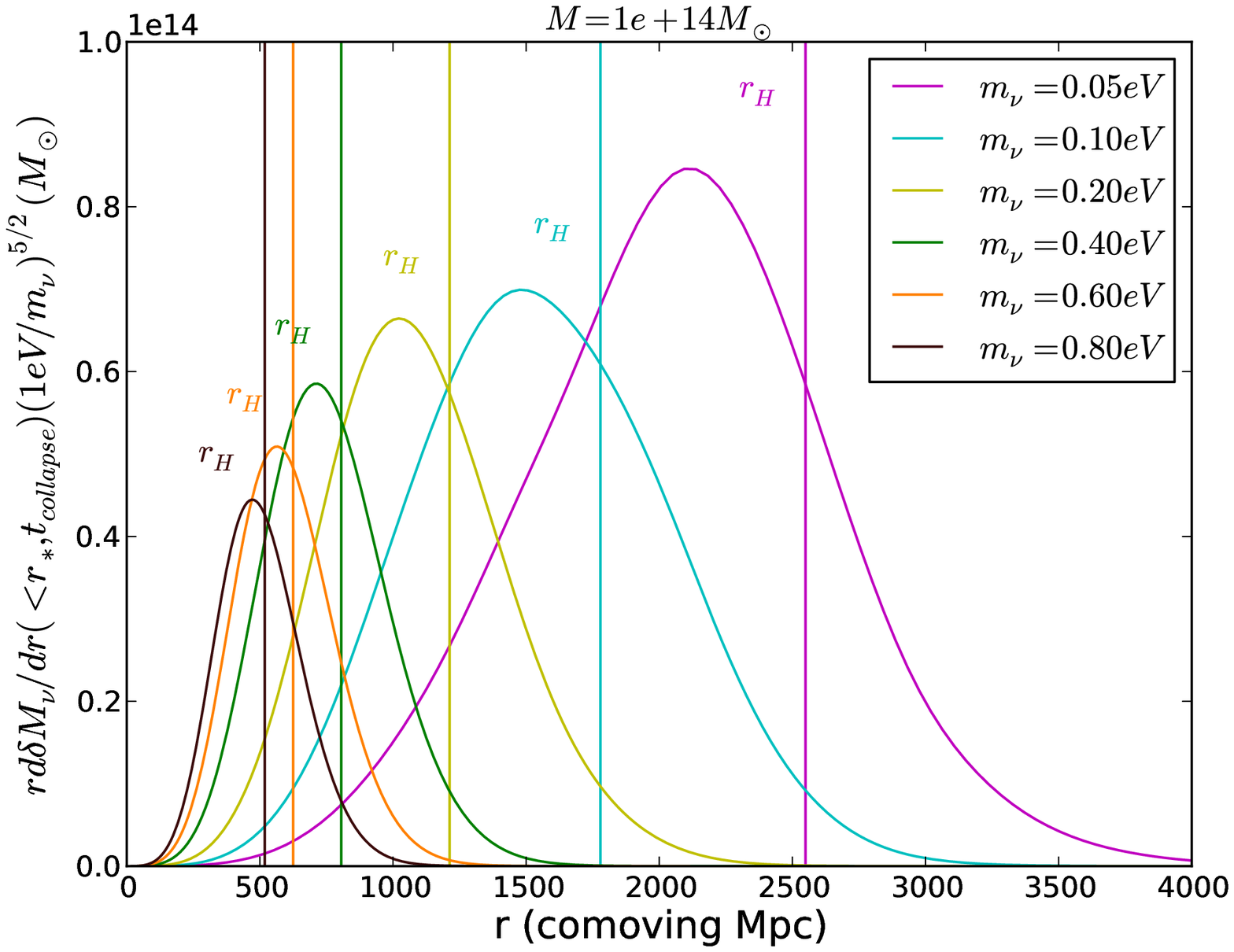} 
\end{tabular}$
\caption{\label{fig:dMnudr} Left: The density profile of neutrino mass from a single neutrino species calculated using the BKT approximation in Eq.~(\ref{eq:Mnuf1}).  Here, the CDM halo has $M = 10^{14} M_\odot$ and collapses around $t_{collapse} \sim$ 8.5 Gyrs, or $z_{collapse} \sim 0.5$. We show several different neutrino masses (the curves roughly have increasing neutrino mass from top to bottom) and scale the results by $m_{\nu}^{3/2}$. The extent of the ``neutrino halo" is large compared to the  CDM virial radius (solid vertical line) and $r_*$ as defined in Eq.~(\ref{eq:rstar}) (dotted vertical lines) appears to be a good characterization of the scale. Right: The neutrinos that contribute to $\delta M_\nu$ originate at a range of distances far from the CDM halo. Plotted is $r d\delta M_\nu/dr(<r_*, t_{collapse})$ scaled by $m_{\nu}^{5/2}$.  The colored vertical lines indicate particle horizon for a neutrino with average momentum $p = 3.151T_\nu$ and corresponding $m_\nu$ (the curves have increasing $m_\nu$ from left to right). }
\end{center}
\end{figure}

 \begin{figure}
\begin{center}$\begin{tabular}{cc}
\includegraphics[width=0.5\textwidth]{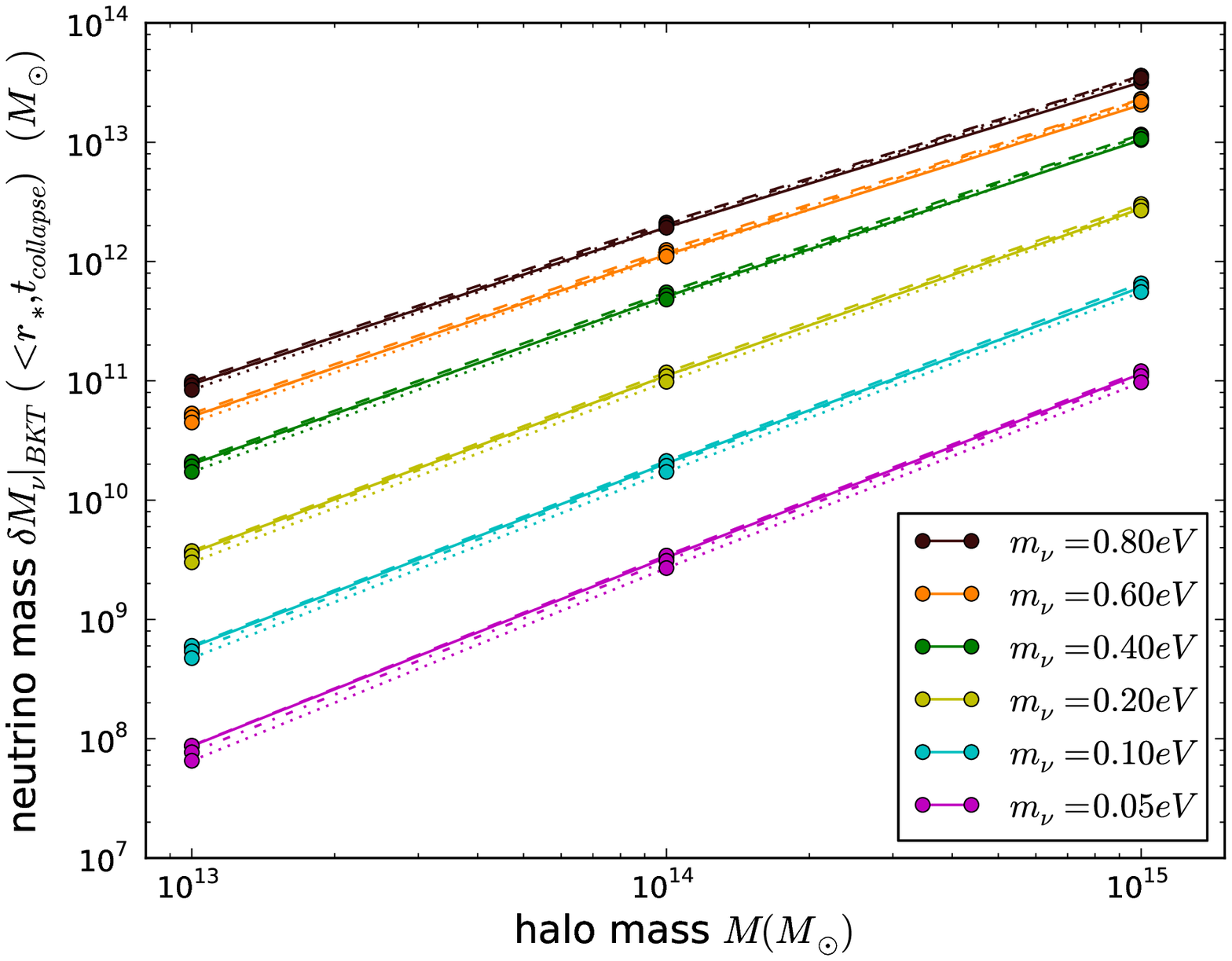} &\includegraphics[width=0.5\textwidth]{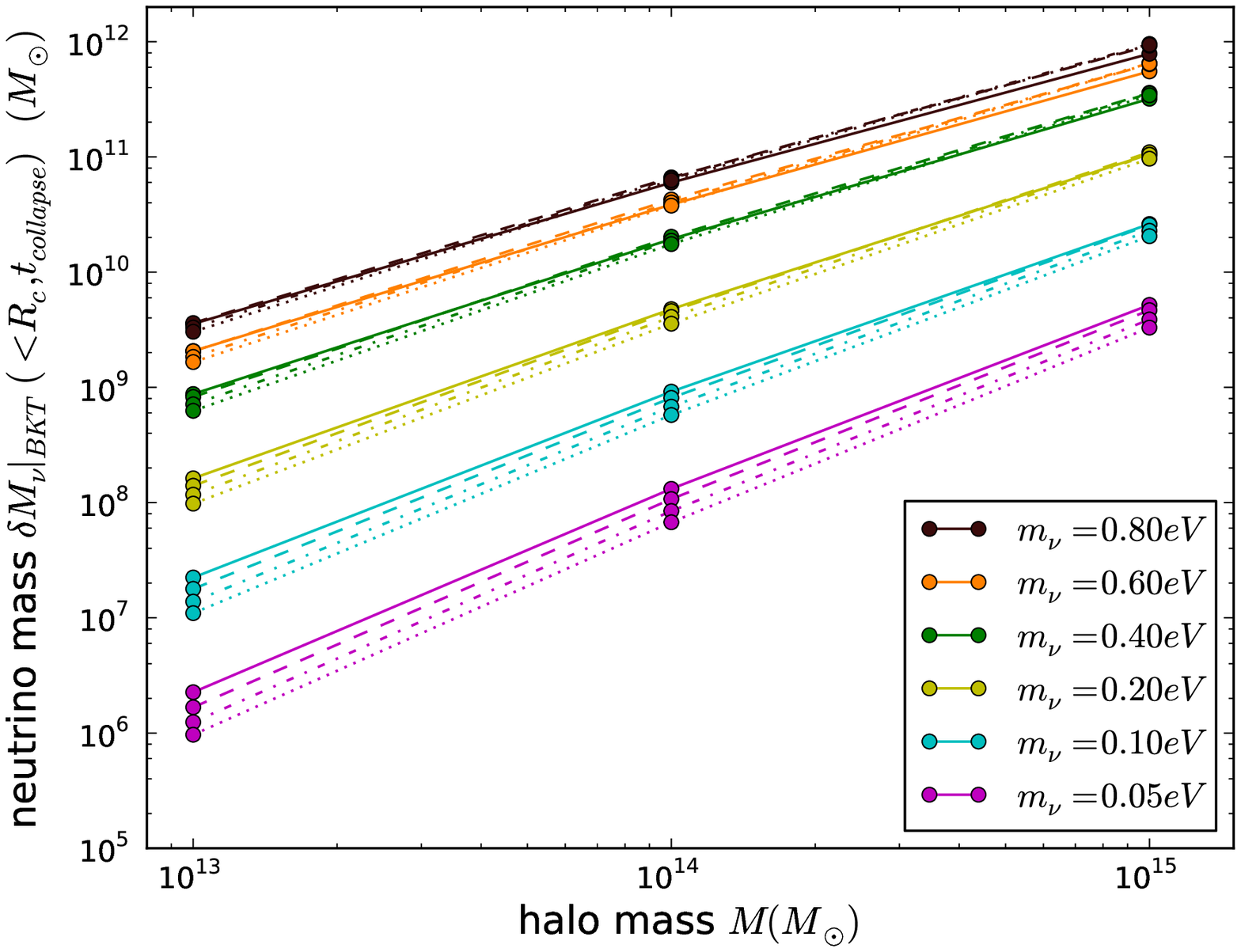} 
\end{tabular}$
\caption{\label{fig:MnuBoltzmann} Left: The neutrino mass fluctuation from a single neutrino species within radius $r_*$ at the collapse time as a function of CDM halo mass calculated using the BKT approximation in Eq.~(\ref{eq:Mnuf1}). The value of $\delta M_\nu$ depends on the halo collapse time and for each ($m_\nu$, $M$) we plot points with a range of $z_{collapse}$ values; they are $z_{collapse} = 0$ (solid), $0.5$ (dashed), $1$ (dot-dashed), and $1.5$ (dotted). Roughly, $\delta M_\nu(<r_*, t_{collapse}) \propto M^{3/2}m_\nu^{5/2}$. Right: The fluctuation in neutrino mass within the CDM halo radius $R_c$ at the collapse time as a function of CDM halo mass, as we show in \S \ref{ssec:full}, the BKT approximation significantly underestimates the mass interior $R_c$ for $m_\nu \gsim 0.1eV$. In both figures the plotted values of $m_\nu$ increase from the bottom curve to the top.}
\end{center}
\end{figure}

In this section we calculate the clustered neutrino mass around a dark matter halo using the approximate solution to the Boltzmann equation for non-relativistic particles from Brandenberger, Kaiser, and Turok (BKT) \cite{Brandenberger:1987kf} (see also \cite{Singh:2002de,Ringwald:2004np, AliHaimoud:2012vj}). Following \cite{Singh:2002de,Ringwald:2004np, AliHaimoud:2012vj}, we change variables in Eq.~(\ref{eq:Boltzmann}) to ${\bf q } = a{\bf p}$ and write the distribution function as a sum of two terms $f({\bf q}, {\bf x}, t) = f_0(q) + f_1({\bf q}, {\bf x}, t)$ where $f_0$ satisfies the homogeneous $\Psi_{pec} =0$ equation. With these changes the Boltzmann equation for non-relativistic neutrinos becomes
\be
\label{eq:boltzmannf1drop}
\frac{\partial f_1}{\partial t} +  \frac{{\bf q}}{a^2m_\nu}\cdot \nabla_x f_1  -  m_\nu \nabla \Psi_{pec}\cdot \nabla_q (f_0 + f_1) =0\,.
\ee
The approximation made in  \cite{Brandenberger:1987kf}  and \cite{Singh:2002de}, which we refer to as the BKT approximation, is to drop the final $\nabla_q f_1$ term.  In this limit Eq.~(\ref{eq:boltzmannf1drop}) is solved by 

 \be
\left.  f_1({\bf x},{\bf q}, t)\right|_{BKT} = -2\frac{m_\nu}{T_\nu} \int_{t_0}^t dt' \frac{ e^{q/T_\nu}}{(e^{q/T_\nu}+1)^2}\hat{q}\cdot \left.\nabla_y\Psi_{pec}({\bf y}, \eta')\right|_{{\bf y} = {\bf x}-{\bf q}(\eta-\eta')/m_\nu}
 \ee
where $\eta$ is a new time variable defined by $a^2d\eta = dt$ so that $q \eta/m_\nu$ is the comoving distance traveled by a (non-relativistic) neutrino along the unperturbed ($\Psi_{pec} = 0$) trajectory. 

For our assumed top-hat density perturbation this gives
\ba
\label{eq:f1}
\left. f_1({\bf x},{\bf q}, t)\right|_{BKT} &=& 2\frac{m_\nu}{T_\nu} \int_{t_0}^t \frac{dt'}{a(t')} \frac{ e^{q/T_\nu}}{(e^{q/T_\nu}+1)^2}\frac{G\delta M(t')}{x^2}\left(\alpha \frac{q}{T_\nu}- \hat{q}\cdot\hat{x}\right)\\
&&\left\{\frac{a^3(t')x^3}{R_c^3}\Theta\left(x^2(1+q^2/T_\nu^2\alpha^2 -2 q/T_\nu\alpha\hat{x}\cdot\hat{q}) < R_c^2(t')/a^2(t')\right) \right.\nn\\
&&\left.+ \frac{\Theta\left(x^2(1+q^2/T_\nu^2\alpha^2 -2 q/T_\nu\alpha\hat{x}\cdot\hat{q}) \ge R_c^2(t')/a^2(t')\right) }{\left(1+q^2/T_\nu^2\alpha^2 -2 q/T_\nu\alpha\hat{x}\cdot\hat{q} \right)^{3/2}}\right\}\nn
\ea

where $\Theta$ is the Heaviside step function and we have defined $\alpha \equiv \frac{T_\nu(\eta-\eta')}{m_\nu |\x|}$. 

With Eq.~(\ref{eq:f1}) in hand, we can calculate the neutrino density perturbation and mass flux at proper position $r$, along with the neutrino mass fluctuation interior to $r$ in the BKT approximation:
\be
\label{eq:deltarhof1}
\left.\delta \rho(r,t) \right|_{{\tiny BKT}}=  m_\nu \int \frac{d^3{\bf q}}{(2\pi)^3}\left.f_1({\bf q}, {\bf r}/a, t)\right|_{BKT}\,,
\ee 

\be
\label{eq:urf1}
\left.\langle u_r\rho_\nu (r, t)\rangle \right|_{{\tiny BKT}}\equiv -\frac{1}{a(t)} \int \frac{d^3{\bf q}}{(2\pi)^3} {\bf q} \cdot \hat{r} \left. f_1({\bf q}, {\bf r}/a(t), t)\right|_{BKT}\,,
\ee

\be
\label{eq:Mnuf1}
\left.\delta M_\nu (< r , t)\right|_{{\tiny BKT}} =  m_\nu \int_{V_r/a^3} d^3 {\bf x} \int \frac{d^3{\bf q}}{(2\pi)^3}\left. f_1({\bf q}, {\bf x}, t)\right|_{BKT}\,.
\ee

Equation (\ref{eq:Mnuf1}) allows us to calculate the neutrino mass profile and check that the radius $r_*$ is a reasonable boundary for the neutrino halo. We can also determine where the neutrino mass that accumulates in the halo at time $t$ originated from at $t\rightarrow 0$. These quantities are shown in Fig.~ \ref{fig:dMnudr}.  We see that $r_*$ is indeed an accurate characterization of the extent of the neutrino perturbation around the halo with radius $R_c$ (see also \cite{VillaescusaNavarro:2012ag}). Note also, that (as we'll show in \S \ref{ssec:full}) at small radii the BKT approximation grows increasingly inaccurate with increasing $m_\nu$. Figure~\ref{fig:dMnudr} also illustrates that neutrinos within $r_*$ originate at a range of distances on either side of the particle horizon. 

The neutrino mass interior to $R_c$ and $r_*$ at the time of halo collapse calculated using the BKT approximation is plotted in Fig.~\ref{fig:MnuBoltzmann}.  We find that for neutrinos with typical velocity fast compared to the escape velocity of the halo the neutrino mass interior to $r_*$ scales roughly as, $\delta M_\nu \propto m_\nu^{5/2} M^{3/2}$, and for neutrinos with typical peculiar velocities slower than the escape velocity, $\delta M_\nu \propto m_\nu^{2} M^{4/3}$. As we show in \S \ref{ssec:full}, the BKT approximation significantly underestimates the mass interior $R_c$ for $m_\nu \gsim 0.1eV$ so the results for $\delta M_\nu(<R_c)$ should be interpreted with caution. 

\subsection{Neutrino capture: ``Absorbing barrier" model of the accretion of bound neutrinos}
\label{ssec:accrete}
 \begin{figure}
\begin{center}$\begin{tabular}{cc}
\includegraphics[width=0.5\textwidth]{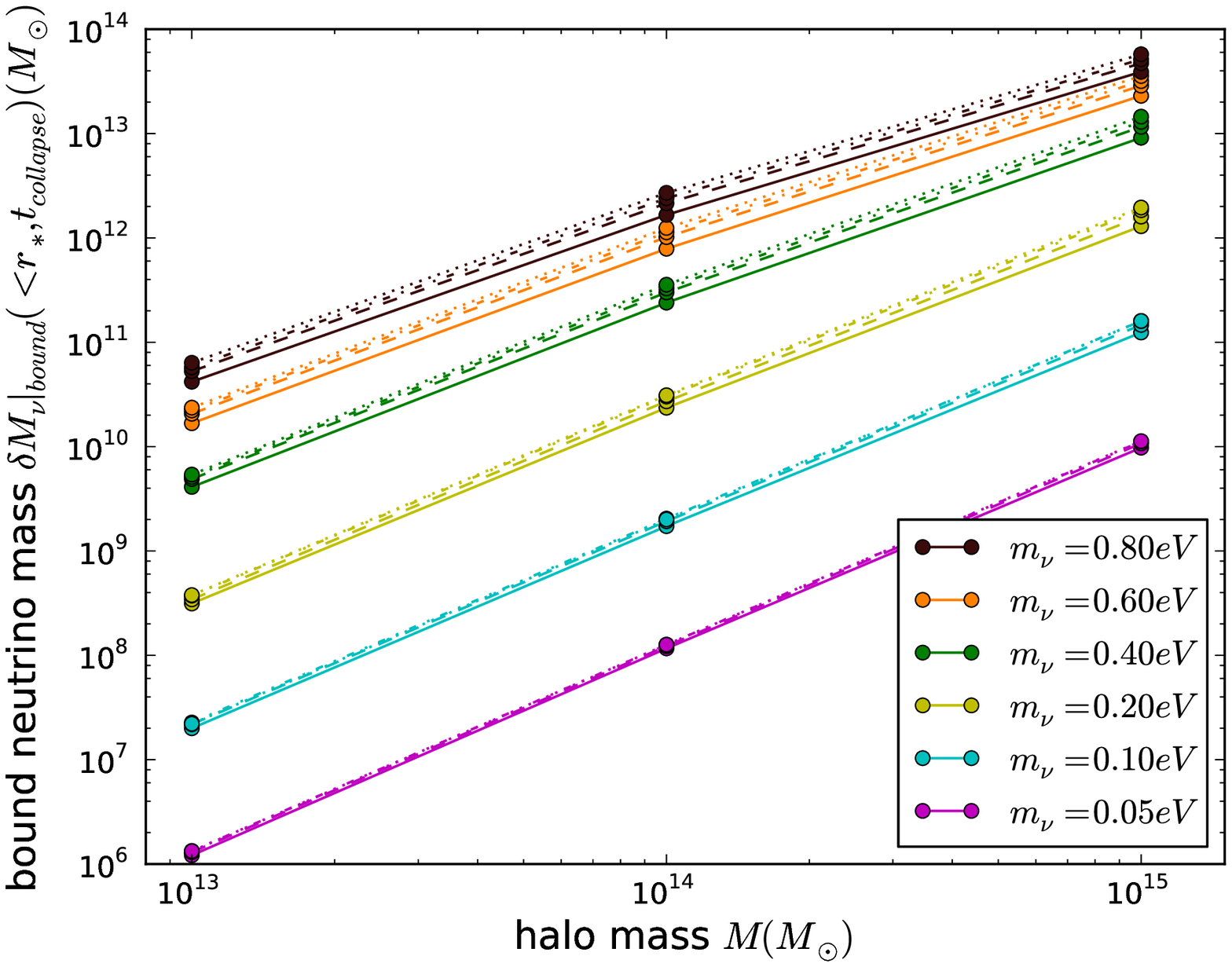}  & \includegraphics[width=0.5\textwidth]{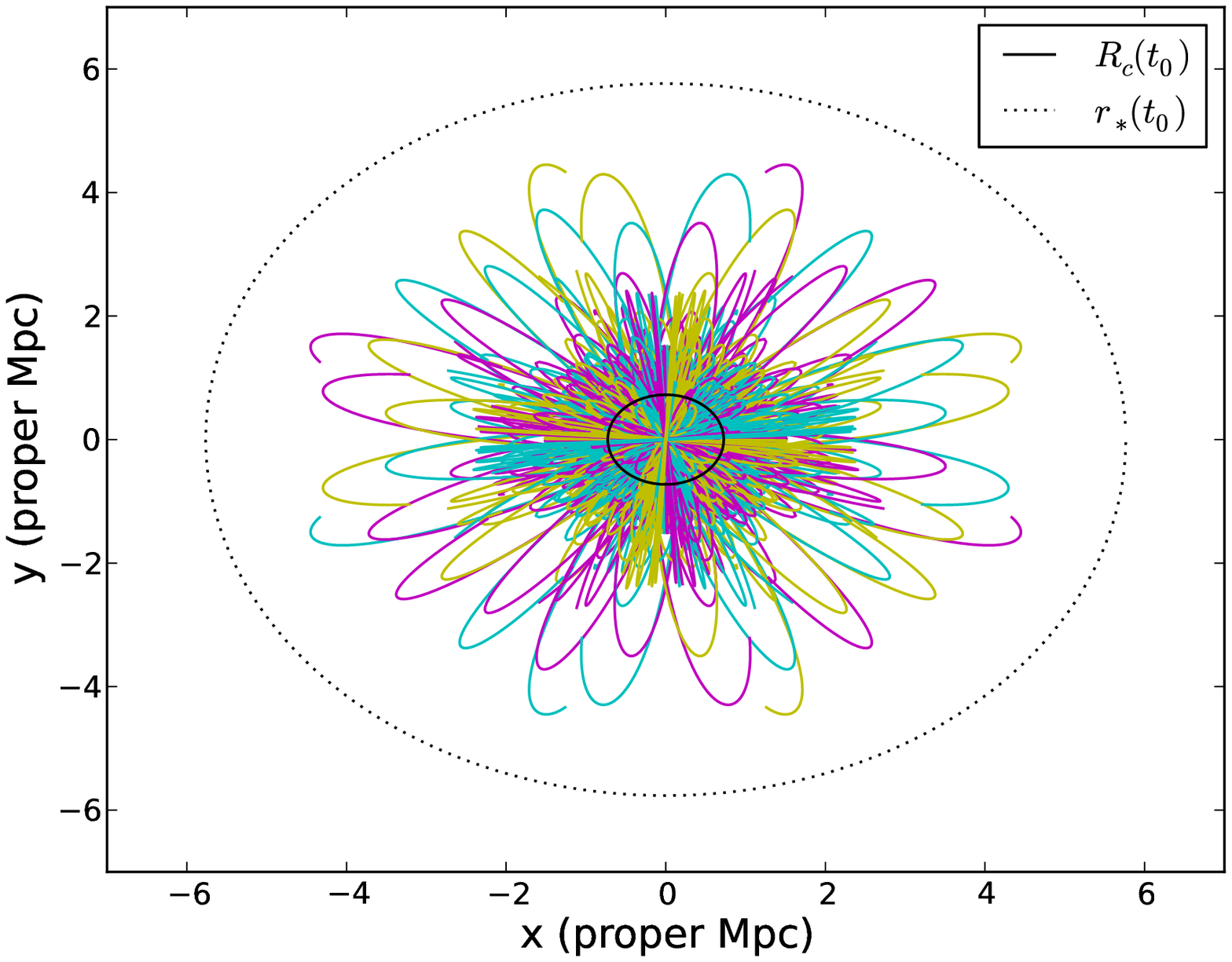}
\end{tabular}$
\caption{\label{fig:MnuCapture} Left: The accreted bound neutrino mass (from a single neutrino species) within radius $r_*$ at the collapse time as a function of CDM halo mass calculated using Eq.~(\ref{eq:dMdt}), Eq.~(\ref{eq:urho}), and Eq.~(\ref{eq:deltarho}). The value of $\delta M_\nu$ depends on the halo collapse time and for each ($m_\nu$, $M$) we plot points with a range of $z_{collapse}$ values;  they are $z_{collapse} = 0$ (solid), $0.5$ (dashed), $1$ (dot-dashed), and $1.5$ (dotted).  For $m_\nu \lsim 0.2 eV$, the bound neutrino mass scales roughly as $\delta M_\nu(<r_*, t_{collapse}) \propto M^{2}  m_\nu^{4} $ whereas for higher neutrino masses the scaling is closer to $\delta M_\nu(<r_*, t_{collapse}) \propto M^{3/2}  m_\nu^{3} $. Right: A subset of the trajectories of $m_{\nu} = 0.05eV$ neutrinos captured by a $M= 10^{14} M_\odot$ halo. Also plotted are the radius of the CDM halo $R_{vir}$ today (solid) and our definition of the boundary of the neutrino halo $r_*$ (dotted) at $z = 0$.  }
\end{center}
\end{figure}

The approximate solution for $f_1$ in \S \ref{ssec:BKT} does not account for neutrinos whose trajectories are significantly perturbed by the halo; in particular it does not properly treat neutrinos that are gravitationally bound to the halo (e.g. the neutrinos with orbiting trajectories such as in Fig.~\ref{fig:rofts}). In this section we develop a model for the rate of accretion of bound neutrinos that we can use to determine the mass in neutrinos that is gravitationally bound to the CDM halo. 

Equation~(\ref{eq:Mnultrstar}) and Eq.~(\ref{eq:Boltzmann}) can be used to get an expression for the neutrino accretion rate at $r_*$, 
\be
\label{eq:dMdt}
\frac{dM_\nu(< r_*)}{d t} = 4\pi r_*^2\left(\frac{dr_*}{dt}-Hr_*\right)\rho(r_*, t) - 4\pi r_*^2 a(t)\int  \frac{d^3{\bf p}}{(2\pi)^3}  {\bf p} \cdot \hat{r}f({\bf x}, {\bf p}, t)
\ee
where we have used the divergence theorem and assumed $f$ vanishes on the momentum boundary, i.e. $f(p=\pm \infty)$.  In principle to calculate $dM_\nu/dt$ from Eq.~(\ref{eq:dMdt})  we need to know the full non-linear distribution function for the neutrinos, $f(\x,{\bf p},t)$. However, we can simplify things considerably by treating the sphere of radius $r_*$ as an absorbing barrier for neutrinos with velocities $|{\bf u}| = |{\bf p}/m_\nu| < u_{esc,\pm}$ where $u_{esc,\pm}$ is the escape velocity of the halo. That is, let $f=0$ for $|{\bf p}/m_\nu|<u_{esc,\pm}$ and ${\bf p}/m_\nu\cdot \hat{r} > dr_*/dt-r_*H$ (outward trajectories) but let neutrinos with $|{\bf p}/m_\nu| > u_{esc}$ leave $r_*$ in the same abundance that they enter so that they do not contribute to $dM_{\nu}/dt$. This approximation is useful because we only need to determine the $f(r_*,{\bf p},t)$ for particles that are entering $r_*$ for the first time. 

Since we have defined $r_*$ as the boundary between the regions where the halo potential dominates and the Hubble flow dominates, we set $f(r_*,p,t) = f_0(p,t)$. Now, in the absorbing barrier approximation the accretion rate of bound neutrino mass can be calculated from
\be
\label{eq:urho}
\left. \langle u_r(r_*,t)\rho_\nu(r_*,t)\rangle\right|_{bound} 
=- \frac{T_\nu  \bar{n}_{1\nu}}{3\zeta(3)a^4} \int_{0}^{1}d\mu \mu  \int_{x_{min}}^{x_{max}} dx\frac{x^3}{e^{| x|}+1}
\ee
\be
\label{eq:deltarho}
\left. \langle \delta\rho_\nu(r_*) \rangle\right|_{bound}  = \frac{m_\nu \bar{n}_{1\nu}}{3\zeta(3)a^3}\int_{0}^1 d\mu  \int_{x_{min}}^{x_{max}} dx\frac{x^2}{e^{ |x|}+1}
\ee
where $x_{max}(\mu, M, t) =m_\nu {\rm min}(dr_*/dt-r_*H_*, u_{esc,+}(\mu) ) a/T_\nu$ and $x_{min} = m_\nu u_{esc,-}(\mu)a/T_\nu$.  Note that Eq.~(\ref{eq:urho}) and Eq.~(\ref{eq:deltarho}) do not make any assumptions about the form (spatial or temporal dependence) of the dark matter halos -- that information, if relevant, goes into determining $u_{\pm}$.  From Eq.~(\ref{eq:urho}) and Eq.~(\ref{eq:deltarho}) we can get a rough estimate of the neutrino accretion rate, 
\ba
\left.\frac{d\delta M_\nu}{dt} \right|_{bound} & = &4\pi r_*^2\left(\frac{dr_*}{dt}-Hr_*\right)\left.\langle \delta\rho_\nu\rangle\right|_{capture} +  4\pi r_*^2 \left.\langle  u_r\rho_\nu\rangle \right|_{capture}\\
& \approx& \frac{\rho_{\nu,massive}  r_*^6H_*^4}{u_{th}^3} 
\ea
which is similar to the usual Bondi accretion formula \cite{Bondi:1952ni} with $c_s=u_{th} = T_\nu/(am_\nu)$. In this limit $d\delta M_\nu/dt \propto m_\nu^4\,\delta M^2$. During matter domination the accretion rate is roughly constant with time in linear regime ($\delta <<1$) and also after virialization. The accretion rate approaches zero during $\Lambda$ domination. 

To determine the accretion rate of bound neutrino mass we find the $u_{esc \pm}$ directly by calculating trajectories of neutrinos with a range of peculiar velocities at $r_*$ and selecting the range of values of $u$ for which neutrinos are bound at the final time (see Appendix \ref{sec:bound} for details). We calculate the net accreted neutrino mass by integrating Eq.~(\ref{eq:urho})-(\ref{eq:deltarho}) with the escape velocities as found numerically in Appendix \ref{sec:bound}. Including a perturbation to the phase space distribution in Eq.~(\ref{eq:dMdt}) calculated from the BKT-approximation changes the final value of  $\left.\delta M_\nu\right|_{bound}$ by $\lsim 25\%$. The $\sim 25\%$ difference between including an $\left. f_1\right|_{BKT}$ term and using $f_0$ only in Eq.~(\ref{eq:dMdt}) occurs for $m_{\nu } = 0.8eV$ and $M = 10^{15}M_\odot$, for masses $\lsim 0.2 eV$ the difference is $\lsim 10\%$ and in all cases it is dominated by the $\langle u_r\rho_\nu\rangle$ term.  Our calculations of the bound neutrino mass do not exceed the Gunn-Tremaine bound \cite{Tremaine:1979we,Kull:1996nx,Ringwald:2004np, Brandbyge:2010ge}. 

Results for the bound accreted neutrino mass are plotted in Fig.~\ref{fig:MnuCapture}. For neutrino masses that are compatible with cosmological bounds ($m_{\nu i} \lsim 0.1-0.2eV$) the bound neutrino mass remains small in comparison with the halo mass ($\lsim1\%$ even for $M = 10^{15} M_\odot$). On the other hand, for the most massive degenerate scenario we consider,  $m_{\nu i } = 0.8 eV$, (which is compatible with terrestrial experiments) the bound neutrino mass interior to $r_*$ can reach $\sim 10\%$ of the total CDM mass interior to $r_*$ and nearly $\sim 20\%$ of the halo mass (the CDM mass within the smaller radius $R_c$). Also plotted in Fig.~\ref{fig:MnuCapture} is a subset of the  neutrino trajectories contributing to $\delta M_{\nu, bound}$: most orbiting neutrinos remain within $(few) \times R_c$, but some trajectories do explore larger radii closer to $r_*$.

\subsection{Full Boltzmann solution}
\label{ssec:full}

 \begin{figure}
\begin{center}$\begin{tabular}{cc}
\includegraphics[width=0.5\textwidth]{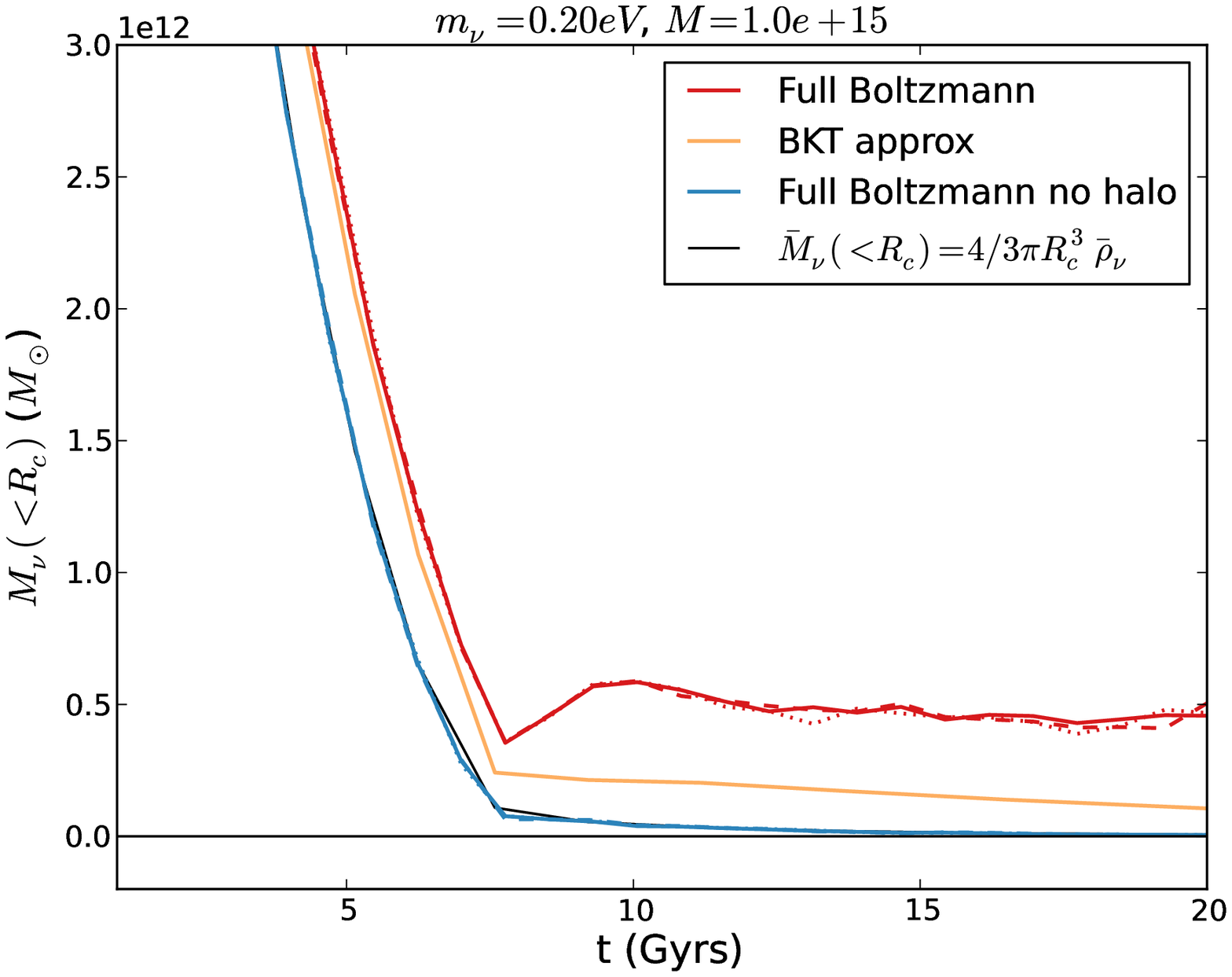} &\includegraphics[width=0.5\textwidth]{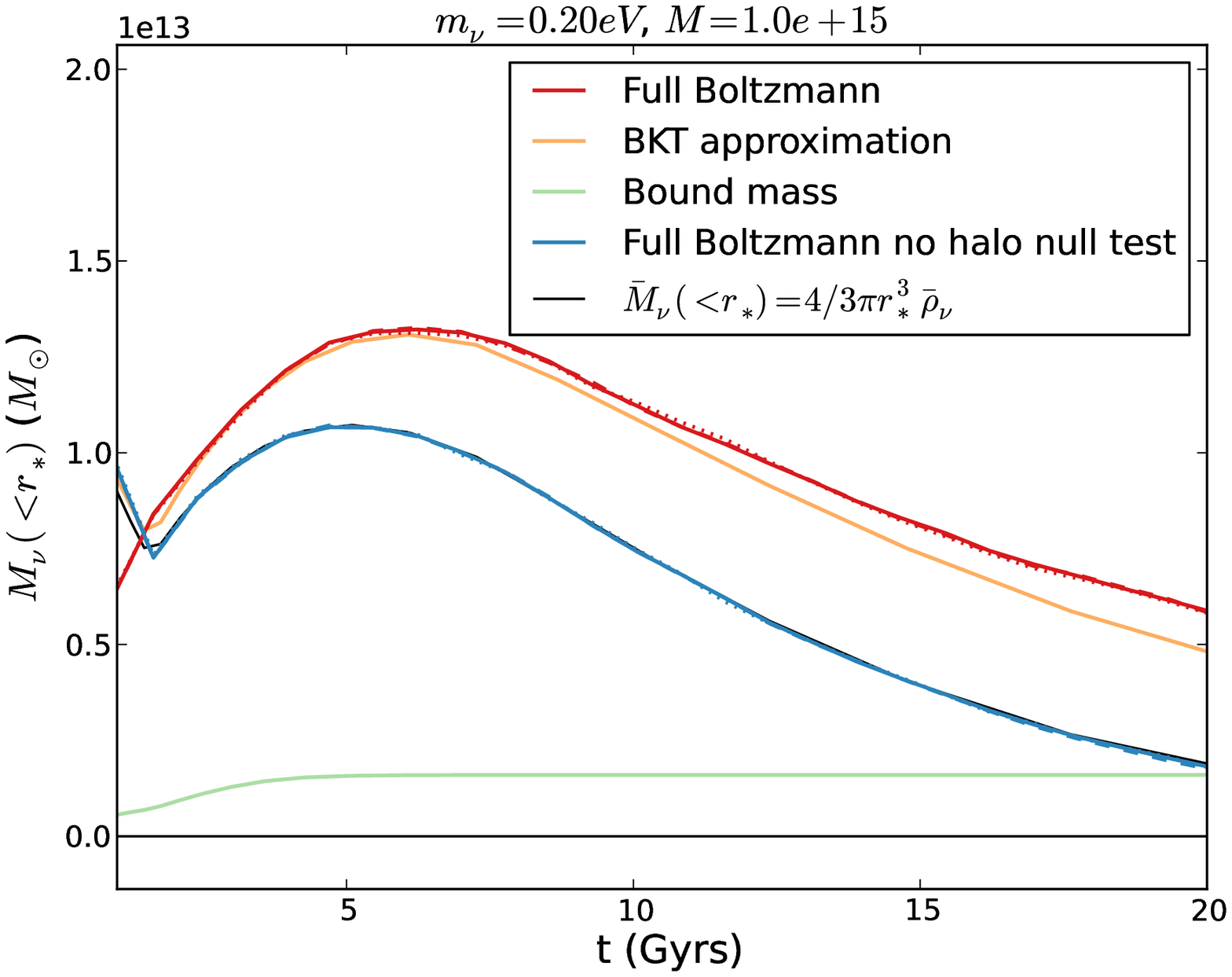} \\
(a)& (b) \\
\includegraphics[width=0.5\textwidth]{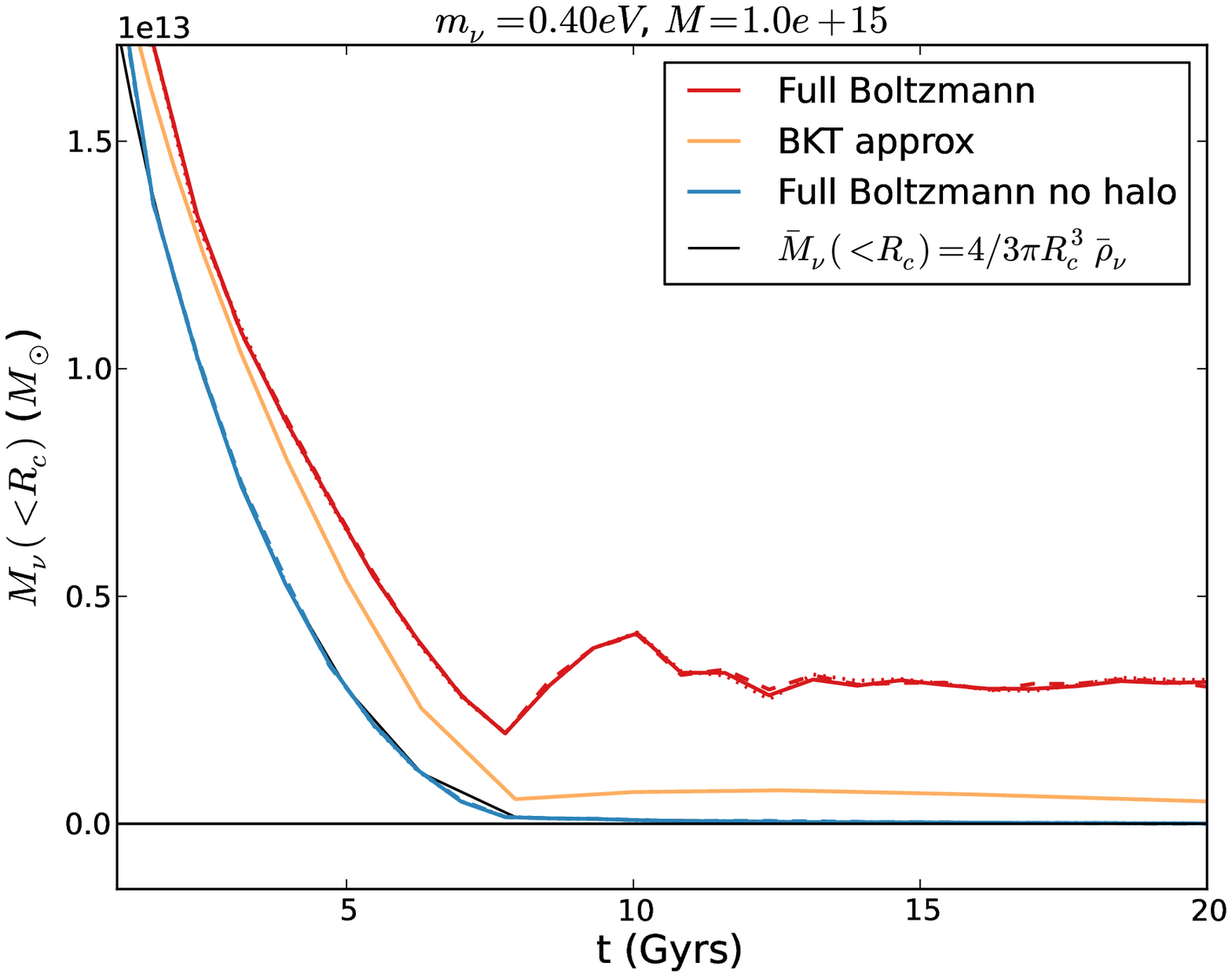} &\includegraphics[width=0.5\textwidth]{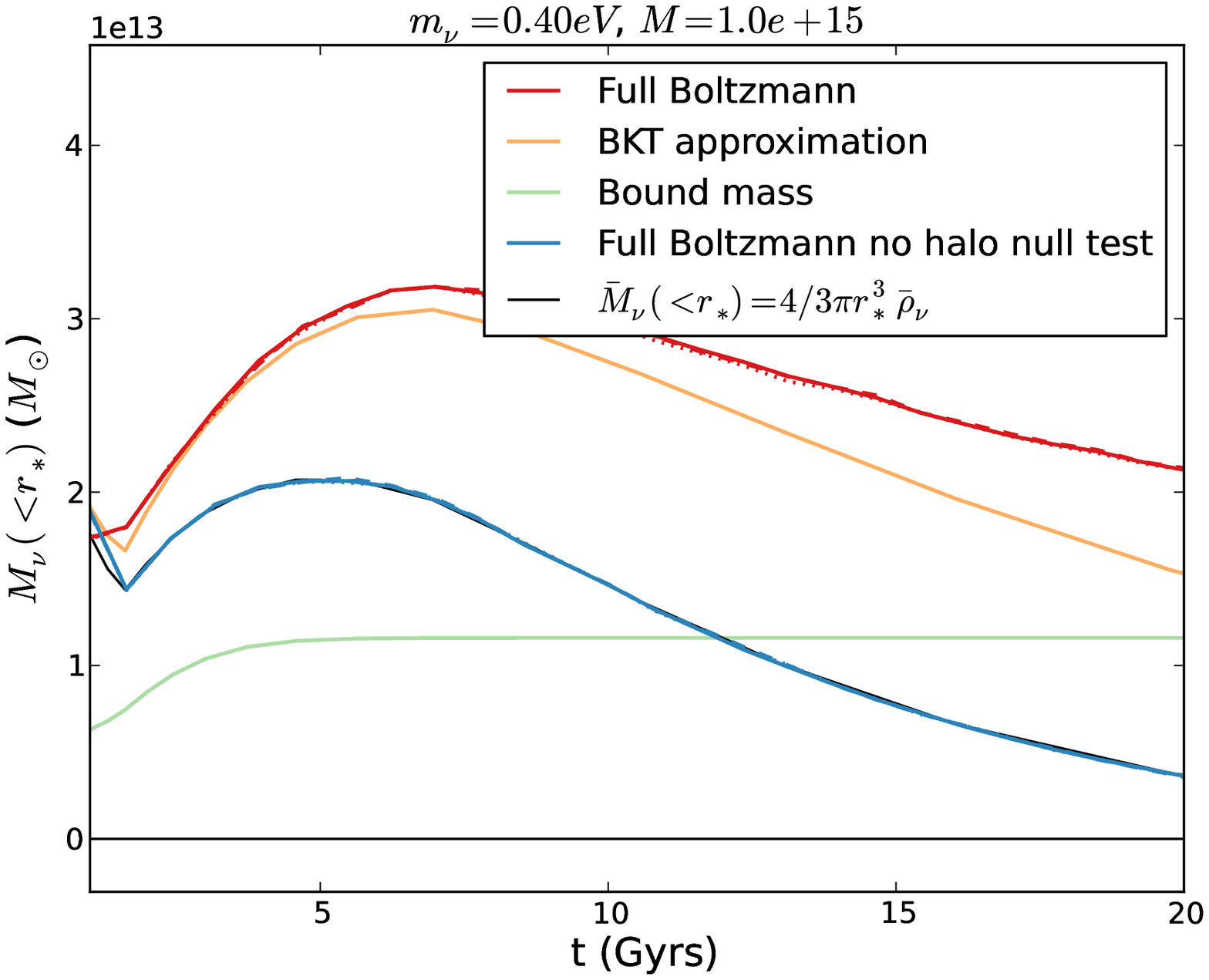} \\
\includegraphics[width=0.5\textwidth]{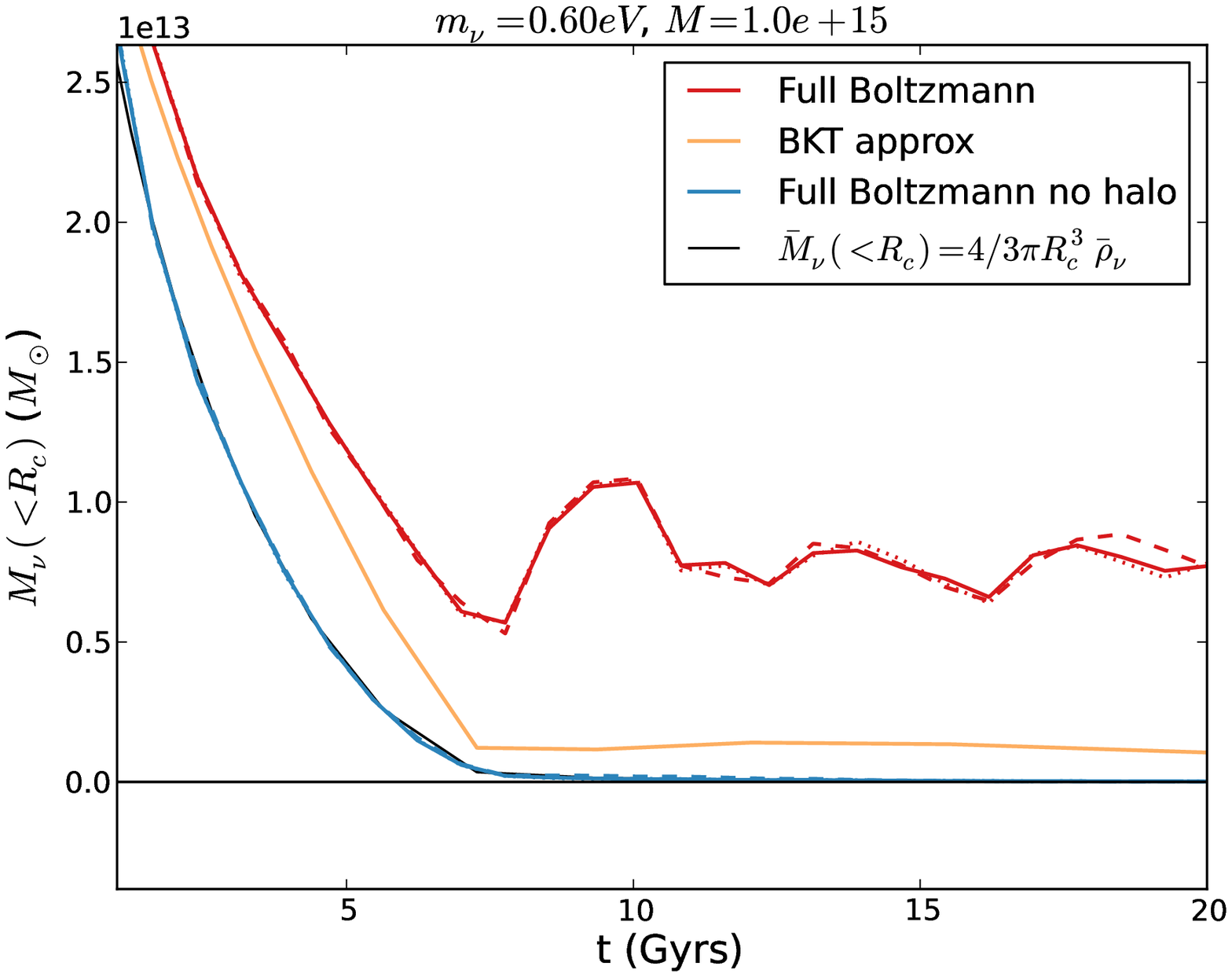} &\includegraphics[width=0.5\textwidth]{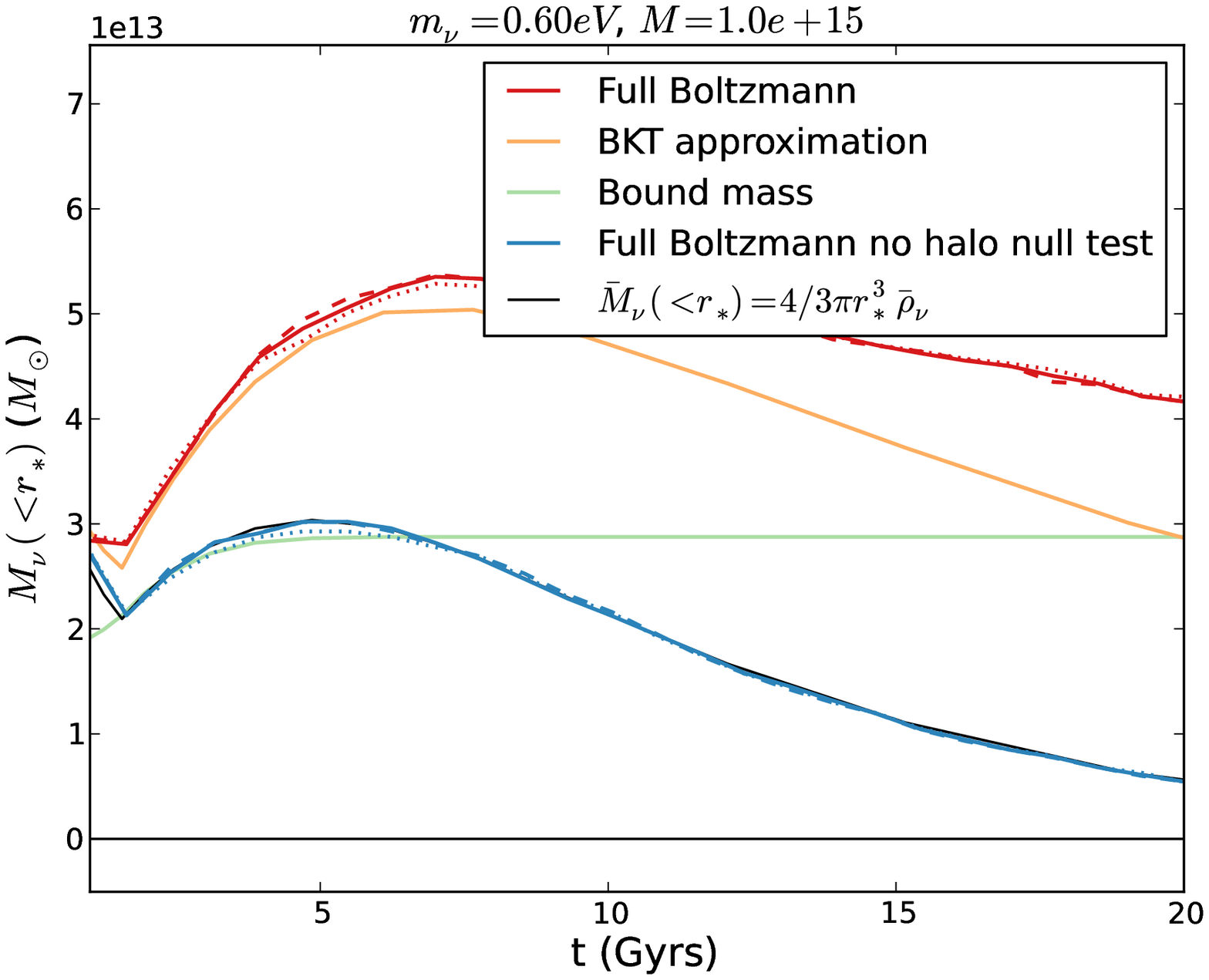} \\
(e)& (f) 
\end{tabular}$
\caption{\label{fig:Mnumcompare}  Left columns: The total neutrino mass interior to the CDM halo radius $R_c$ calculated using the approximate solution to the Boltzmann equation in \S\ref{ssec:BKT} (yellow) and the full Boltzmann solution (red curves). Right columns: The total neutrino mass interior to $r_*$ (see Eq.~(\ref{eq:rstar})) calculated using the approximate solution to the Boltzmann equation in \S\ref{ssec:BKT} (yellow) and the full Boltzmann solution (red curves). Also plotted is the accreted bound neutrino mass calculated using the absorbing barrier model given in \S \ref{ssec:accrete} (green curves).  In each panel $M = 10^{15} M_\odot$ and the halo collapses at $z_{collapse}\sim 0.5$, or $t\sim 8.5 Gyrs$. Each row shows the clustering of a single neutrino species with mass $m_\nu = $ $0.20 eV$ (top), $ 0.40 eV$ (middle), and  $0.60 eV$ (lower). The blue curves are tests of the full Boltzmann code for the mass interior to $R_c$ or $r_*$ assuming no halo is present. We also plot some convergence tests: the dashed and dotted curves use our full Boltzmann calculation with half the number of points in initial comoving radial position (dashed) and peculiar velocity magnitude (dotted).}
\end{center}
\end{figure}

The approximate solution in \S \ref{ssec:BKT} does not accurately treat the bound neutrinos. As we have seen in \S \ref{ssec:accrete}, bound neutrinos can be a significant contribution to the total $\delta M_\nu$ around the halo. In this section we sample the initial neutrino phase space and numerically integrate the trajectories in the external halo potential to determine the neutrino clustering (similar to the ``N-1-body" approach of \cite{Ringwald:2004np,Brandbyge:2010ge}). Here, our only approximation is to assume that the change to the CDM halo potential due to neutrino clustering can be ignored when calculating the neutrino trajectories. 

Precisely, our method here is as follows. At $t = 1.5 t_{NR}$, where $t_{NR}$ is defined as the time at which $p/m_\nu < 0.05$, we numerically integrate Eq.~(\ref{eq:NREOM}) with initial positions and momenta taking values on a uniformly spaced grid in $r_i$, $p_i$. The ranges of $r_i$ and $p_i$ are determined by the unperturbed neutrino distribution function in Eq.~(\ref{eq:f0}) and the range of positions between the halo origin and the maximum travel distance between $1.5t_{NR}$ and $t$ for the bin with highest initial momentum. From this grid of trajectories we can compute properties of the neutrino distribution at later times by numerically integrating over the volume of initial phase space that satisfies our criterion (e.g. the trajectories from that volume that are within $r_*$ at time $t$). For the neutrino mass interior to $r_*$ we have, 
\be
M_\nu(<r_*, t)  = \int d^3{\bf r}_i\int\frac{d^3{\bf p}_i}{(2\pi)^3}f_0(p_i) \Theta\left(|{\bf r}(t | {\bf r}_i, {\bf p}_i)| < r_* \right)\,.
\ee
This calculation is computationally intensive and for smaller halo masses and smaller neutrino masses it is increasingly difficult to achieve sufficient sampling of the initial phase space for convergence. Fortunately these are precisely the scenarios in which we expect the BKT approximation to be accurate. As a test of our calculations we use the same method to determine the neutrino mass interior to $r_*$, $R_c$ in the absence of the halo potential, that is, if our calculation has converged we should recover $\bar{M}_\nu(<r_*) = 4/3\pi r_*^3\bar{\rho}_{\nu}$,  $\bar{M}_\nu(<R_c) = 4/3\pi R_c^3\bar{\rho}_{\nu}$.  

In comparing this ``full Boltzmann" calculation to the BKT approximation in \S \ref{ssec:BKT} we find that for $m_{\nu } \lsim 0.2eV$ and $M \lsim 10^{14} M_\odot$, the BKT approximation for $\delta M_\nu (<r_*)$ is accurate to about $\sim 10\%$. In Fig.~\ref{fig:Mnumcompare} results for $m_{\nu} = 0.2eV$, $m_{\nu} = 0.4eV$ and $m_\nu = 0.6eV$ are plotted.  For larger neutrino masses and larger halo masses, the BKT approximation can underestimate the neutrino mass within $r_*$, but in no case that we consider is $M_{\nu}$ off by more than $\sim 50\%$ today (even for $m_{\nu } = 0.8eV$ and $M = 10^{15} M_\odot$ the error in $\delta M_\nu(<r_*)$ is $\sim 50\%$). In the extreme $\Lambda$-dominated future the BKT approximation is worse: the true $\delta M_{\nu} (<r_*)$ approaches a constant while in the BKT calculation $\delta M_{\nu}$ continues to fall.  As noted by others \cite{Ringwald:2004np,Brandbyge:2010ge}, the BKT approximation underestimates the neutrino mass fluctuation on the smaller scale of the halo radius $R_c$ by a large amount. The BKT approximation underestimates $\delta M_\nu(<R_c)$ by nearly a factor of $3$ for $m_{\nu} = 0.4eV$ and an order of magnitude for $m_{\nu} = 0.8eV$. However, for values of the neutrino mass that are within the current cosmological bounds, $m_{\nu} \lsim 0.2eV$ say, the BKT approximation is accurate to about a factor of $2$ even on the scale $R_c$.

\section{Results for the neutrino mass around spherical halos}
\label{sec:totalmass}

\begin{figure}
\begin{center}$\begin{tabular}{cc}
\includegraphics[width=0.5\textwidth]{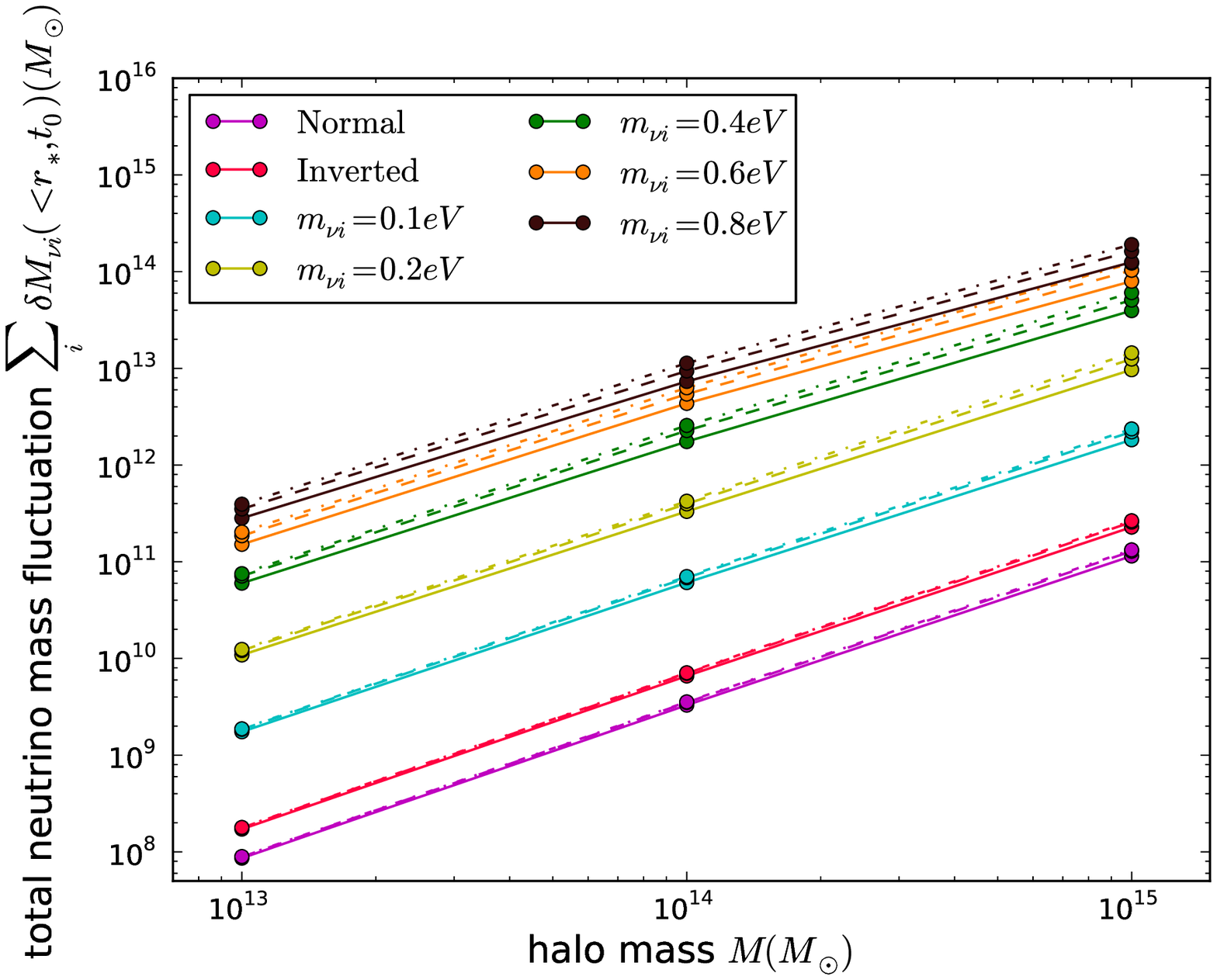} &\includegraphics[width=0.5\textwidth]{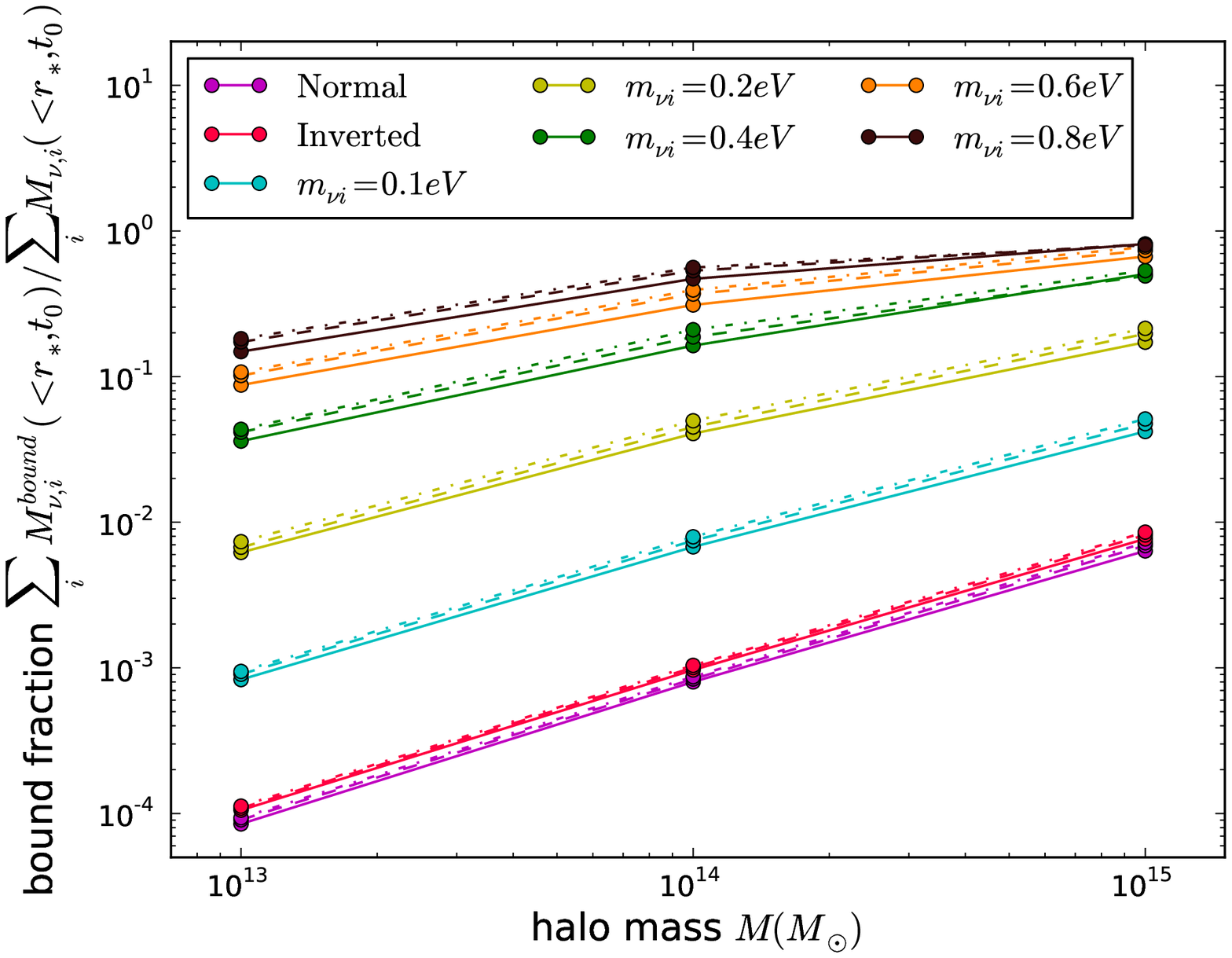} 
\end{tabular}$
\caption{\label{fig:MnuvsM} Left: The total fluctuation in neutrino mass interior to $r_*$ at $z =0$. Shown is $\delta M_{\nu}(<r_*)$ as a function of halo mass $M$ for several neutrino mass hierarchy scenarios indicated by different color lines. From bottom top they are: normal hierarchy ($m_{\nu 1} = 0.05 eV$, $m_{\nu 2} = 0.01eV$, and $m_{\nu 3} = 0eV$), inverted hierarchy  ($m_{\nu 1} = 0.05 eV$, $m_{\nu 2} = 0.05eV$, and $m_{\nu 3} = 0eV$), degenerate $m_{\nu i} = 0.1eV$, degenerate $m_{\nu i} = 0.2eV$,  degenerate $m_{\nu i} = 0.4eV$ , degenerate $m_{\nu i} = 0.6eV$,  degenerate $m_{\nu i} = 0.8eV$). At a fixed redshift (above $z\sim 0$), the neutrino mass fluctuation in a given halo depends on the time of halo collapse, shown above are $z_{collapse} \sim 1$ (dot-dashed), $z_{collapse} \sim 0.5$ (dashed), and $z_{collapse} \sim 0$ (solid). Right: The fraction of the total neutrino mass interior to $r_*$ that is bound to the halo. }
\end{center}
\end{figure}

In Fig. \ref{fig:MnuvsM} we plot our final results for the neutrino mass interior to $r_*$, our definition of the boundary of the neutrino halo, and the fraction of that mass that is bound to the halo at late times. To calculate the total neutrino mass, we use the BKT approximation of \S \ref{ssec:BKT} for $m_{\nu} \le 0.1 eV$ , while for $m_{\nu } = 0.2eV$, we use the BKT approximation for $M = 10^{13}M_\odot$, $10^{14} M_\odot$ but the exact Boltzmann calculation of \S \ref{ssec:full} for $M = 10^{15}$, and for $m_{\nu} \ge 0.4eV$ we use the BKT only for $M_{halo } = 10^{13} M_\odot$ and the full Boltzmann calculation in all other cases.  For the accreted bound mass we use Eq.~(\ref{eq:dMdt}) with an unperturbed Boltzmann distribution for $m_{\nu} \le 0.1 eV$, but include a perturbation calculated from the BKT approximation in Eq.~(\ref{eq:f1}) for $m_{\nu } = 0.2eV$ and $M = 10^{15}M_\odot$, as well as for $m_{\nu} \ge 0.4eV$ when $M = 10^{14}M_\odot$, $10^{15}M_\odot$.   We calculate the neutrino mass fluctuation interior to CDM halos with a range of halo masses and collapse times.  For neutrinos with masses $m_{\nu i}\lsim 0.2eV$, $\delta M_\nu(<r_*)$ does not vary strongly with redshift and our calculations of the neutrino mass within $r_*$ today are well approximated by
\be
\label{eq:Mnurstarfit}
\delta M_{\nu}(<r_*, t_0)  \approx  \sum_{i} \left(3.4 \times 10^{9} M_\odot\right)\left(\frac{m_{\nu i}}{0.05 eV}\right)^{2.6}\left(\frac{M}{10^{14} M_\odot}\right)^{1.5}
\ee
where $M$ is the mass of CDM. For the same mass range, the bound neutrino mass today is well approximated by
\be
\label{eq:Mnuboundfit}
\left.  \delta M_{\nu}(<r_*, t_0) \right|_{bound} \approx  \sum_{i} \left(1.2 \times 10^{8} M_\odot\right)\left(\frac{m_{\nu i}}{0.05 eV}\right)^{3.8}\left(\frac{M}{10^{14} M_\odot}\right)^{1.9}\,.
\ee
For larger neutrino masses, both  $\delta M_{\nu}(<r_*, t_0)$ and  $\left.\delta M_{\nu}(<r_*, t_0)\right|_{bound}$ depend more strongly on the redshift of halo collapse and, the dependence on $m_\nu$ and $M$ is more complicated than the product of power laws given above. For instance, for $m_{\nu}\gsim 0.4eV$, $\delta M_{\nu}(<r_*, t_0)$ varies by a factor of $\mathcal{O}(1)$ between halos that collapse at $z\sim 0$ and $z\sim 1.5$, with the larger changes occurring for high mass halos and larger neutrino masses.

\section{Conclusion} 
\label{sec:conclusion}

We have investigated neutrino clustering in the simplest model of halo formation: the spherical collapse model for an isolated halo. The methods of analysis here can straightforwardly be applied to more realistic models of dark matter halos if the form of the halo potential is given. However, even in this simple model the neutrino halos are comparatively more complicated than the dark matter and there are several interesting takeaway lessons. First, the physical extent of the neutrino halo is significantly larger than the virial radius of the dark matter halo -- a factor of $\sim 8$ for a virialized halo during matter domination. Despite the fact that the neutrino mass contributes only a small fraction to the total mass of the halo, the neutrino mass is more spatially extended and, in this simple model, the neutrino density perturbation dominates over CDM density perturbation at large distances. While it would be extremely challenging to detect the neutrino halo, it is at least in principle possible with weak gravitational lensing \cite{VillaescusaNavarro:2011ys}. This result for spherical halos is in qualitative agreement with the results of \cite{VillaescusaNavarro:2012ag}, who found that at large radii the neutrino density profile around CDM halos in their simulations can be fit by $\delta\rho_\nu \propto r^{-\alpha}$ with $\alpha \sim 1$, whereas CDM at those distances follows an NFW profile with $\delta \rho_c\propto r^{-3}$.  Another point is that the total neutrino mass that remains bound around the halo is a weak function of the halo collapse time with halos that collapse earlier accumulating neutrino mass (see Fig.~\ref{fig:MnuvsM}). 

For neutrino masses that are well within the cosmological bounds ($m_\nu \lsim 0.1eV$, say) the total neutrino mass and bound neutrino mass interior to $r_*$ does not vary too much with the halo collapse time. We have provided power law fitting formulae for both $\delta M_{\nu}(<r_*)$ and $\left. \delta M_\nu(<r_*)\right|_{bound}$ in Eq.~(\ref{eq:Mnurstarfit}) and Eq.~(\ref{eq:Mnuboundfit}) that are accurate to $\sim 20\%$ and $\sim 35\%$ respectively for $m_{\nu,i} \lsim 0.2eV$ across the range of halo masses and collapse redshifts that we have considered in Fig.~\ref{fig:MnuvsM}. While $r_*$ is a more appropriate characterization of the radius of the neutrino halo the neutrino mass interior to the virial radius of CDM may be of interest as well. We caution that to determine $\delta M_{\nu}(<R_c)$ to within $\mathcal{O}(1)$, solving the full Boltzmann equation is necessary (see Fig.~\ref{fig:Mnumcompare}). 

In this paper we have considered isolated halos that are at rest with respect to the cosmological frame. However, for a halo in a network of large scale structure there may be important changes.  First, for a halo in the cosmic web the gravitational effects of nearby structure may truncate the neutrino halo at a radius smaller than our quoted $r_*$ (which was set by the scale where cosmic expansion becomes important). Additionally, neutrinos that cluster around a halo originate at great distances from the halo itself (of order $H^{-1}_0$, see Fig.~\ref{fig:dMnudr}), therefore the distribution of neutrinos reaching the halo should be nearly isotropic in the cosmological frame  -- as we have calculated here (isotropy of neutrino velocity field around the halo is also seen in the simulations of \cite{VillaescusaNavarro:2012ag}). On the other hand, the cold dark matter that composes most of the halo mass comes from just a few comoving $Mpc$ around the halo and can therefore have large angular momentum and bulk velocity in comparison with that of the neutrinos. A halo moving with a bulk flow, will then see a dipolar distribution of neutrinos, leading to a relative velocity effect for neutrinos and CDM akin to the baryonic relative velocity effect of \cite{Tseliakhovich:2010bj}. A study of bulk motions on neutrino clustering will be presented elsewhere \cite{LoVerde2014}. 

\acknowledgements
M.L. is grateful for discussions with Yacine Ali-Ha\"{i}moud, Neal Dalal, Chris Hirata, Wayne Hu, and Andrey Kravtsov. M.L. is grateful for hospitality at the Institute for Advanced Study while this work was being completed. M.L. is supported by U.S. Dept. of Energy Contract No. DE-FG02-90ER-40560. M.Z. is supported in part by the National Science Foundation Grants No. PHY- 0855425, No. AST-0907969, and No. PHY-1213563

\appendix
\section{Geodesic equation for massive neutrinos and the Newtonian limit}
\label{sec:GRNewton}
 Let's start by considering the trajectory of a single neutrino in an expanding Universe with a spherical CDM density perturbation described by potentials $\Phi$ and $\Psi$.  The metric is
\be
ds^2=-(1+2\Psi(\rc,\tc))d\tc^2+a^2(\tc)(1+2\Phi(\rc,\tc))(d\rc^2+\rc^2d\Omega^2)
\ee
where the underbars distinguish comoving coordinates $(\rc,\tc)$ from proper coordinates $(r,t)$ used in the rest of the paper. 

The momentum measured by a comoving observer is $p^i=aE(1+\Phi-\Psi)d\b {\it x}^i/d\tc$ where $E^2=m^2+p^2$. 
The time component of the geodesic equation gives the evolution of the energy ($d\tc/d\lambda=(1-\Psi)E$). For an initially radial path, this becomes
\be
\label{eq:dEdtradial}
\frac{1}{E}\frac{dE}{d\tc} +\partial_{\rc}\Psi \frac{d\rc}{d\tc}+a^2\left(\mathcal{H}(1+2\Phi-2\Psi)+{\Phi}'\right)\left(\frac{d\rc}{d\tc}\right)^2=0\,. 
\ee
or 
\be
\label{eq:dpdt}
\frac{d p}{dt}=-p\mathcal{H}-p{\Phi}'-\frac{E}{a}\hat{p}^i\partial_{i}\Psi
\ee
where $' = \partial/\partial\tc$ and $\mathcal{H}=a'/a$. For a particle on a radial trajectory the spatial geodesic equation in combination with Eq.~(\ref{eq:dEdtradial}) gives, 
\be
\frac{d^2\rc}{d\tc^2}+2(\mathcal{H}+\Phi')\left(1-\frac{a^2}{2}(1+2(\Phi-\Psi))\left(\frac{d\rc}{d\tc}\right)^2\right)\frac{d\rc}{d\tc}-\Psi'\frac{d\rc}{d\tc}+\frac{1}{a^2}\partial_{\rc}\Psi\left(1-2a^2\left(\frac{d\rc}{d\tc}\right)^2\right)+\partial_{\rc}\Phi\left(\frac{d\rc}{d\tc}\right)^2=0\,.
\ee
Defining the peculiar velocity $u=a(1+\Phi-\Psi)\frac{d\rc}{d\tc}$ and $\gamma=1/\sqrt{1-u^2}$, the above can be rewritten in a more compact form, 
\be
\label{eq:RelEOM}
\frac{d}{d\tc}\left(\gamma u\right)=-\gamma\left((\mathcal{H}+\Phi')u+\frac{1}{a}\partial_{\rc}\Psi\right)\,.
\ee
which is the equation of motion for a relativistic point particle with a friction term $-\gamma(\mathcal{H}+\Phi')u$. 

Given $\Phi(\rc,\tc)$, the initial position and velocity, the above can be integrated numerically to give particle trajectories. Notice that the particle mass doesn't appear in Eq.~(\ref{eq:RelEOM}), so the only difference for relativistic and non-relativistic particles is in the initial conditions for $d\rc/d\tc$. We found that it is numerically more stable to solve for $p$ from Eq.~(\ref{eq:dpdt}) and $d\rc/d\tc=p/aE(1-\Phi+\Psi)$, which includes the mass dependence and forces $-m^2=p^2-E^2$ at each time step.

To make contact with the Newtonian equation of motion given in Eq.~(\ref{eq:NREOM}) we first take the non-relativistic limit of Eq.~(\ref{eq:RelEOM}) (i.e. drop terms $\mathcal{O}(u^2)$)
\be
\label{eq:NRu}
\frac{du}{d\tc} =-(\mathcal{H}+\Phi')u-\frac{1}{a}\partial_{\rc}\Psi\,.
\ee
Now define the proper distance ${\bf r}=a(1+\Phi){\bf \rc}$ and proper time $dt=(1+\Psi)d\tc$. The proper velocity is then 
\be
\label{eq:vdefine}
v\equiv \frac{d r}{dt}=Hr+\dot\Phi r+u +r^2H\partial_r\Phi+ur\partial_r\Phi
\ee
where $\dot{}=\partial/\partial t$, the derivative with respect to proper time at fixed proper distance (related to the time derivative at fixed comoving distance through $\partial_{\tc} = \partial_t-rH\partial_r$). Substituting
Eq.~(\ref{eq:vdefine}) into Eq.~(\ref{eq:NRu}) gives
\be
\frac{d v}{dt}=\left(\dot{H}+\ddot{\Phi}+\left(H+\dot\Phi\right)^2\right)r-\partial_r\Psi+\mathcal{O}(r^2H^2)\,.
\ee
The $\mathcal{O}(r^2H^2)$ terms can be safely neglected for dynamics on scales small compared to the horizon. Further dropping the $\dot\Phi$ and $\ddot\Phi$ terms 
gives
\be
\frac{d v}{dt}=\left(\dot{H}+H^2\right)r-\partial_r\Psi
\ee
as stated in Eq.~(\ref{eq:NREOM}). 

\begin{figure}
\begin{center}$\begin{array}{cc}
\includegraphics[width=0.5\textwidth]{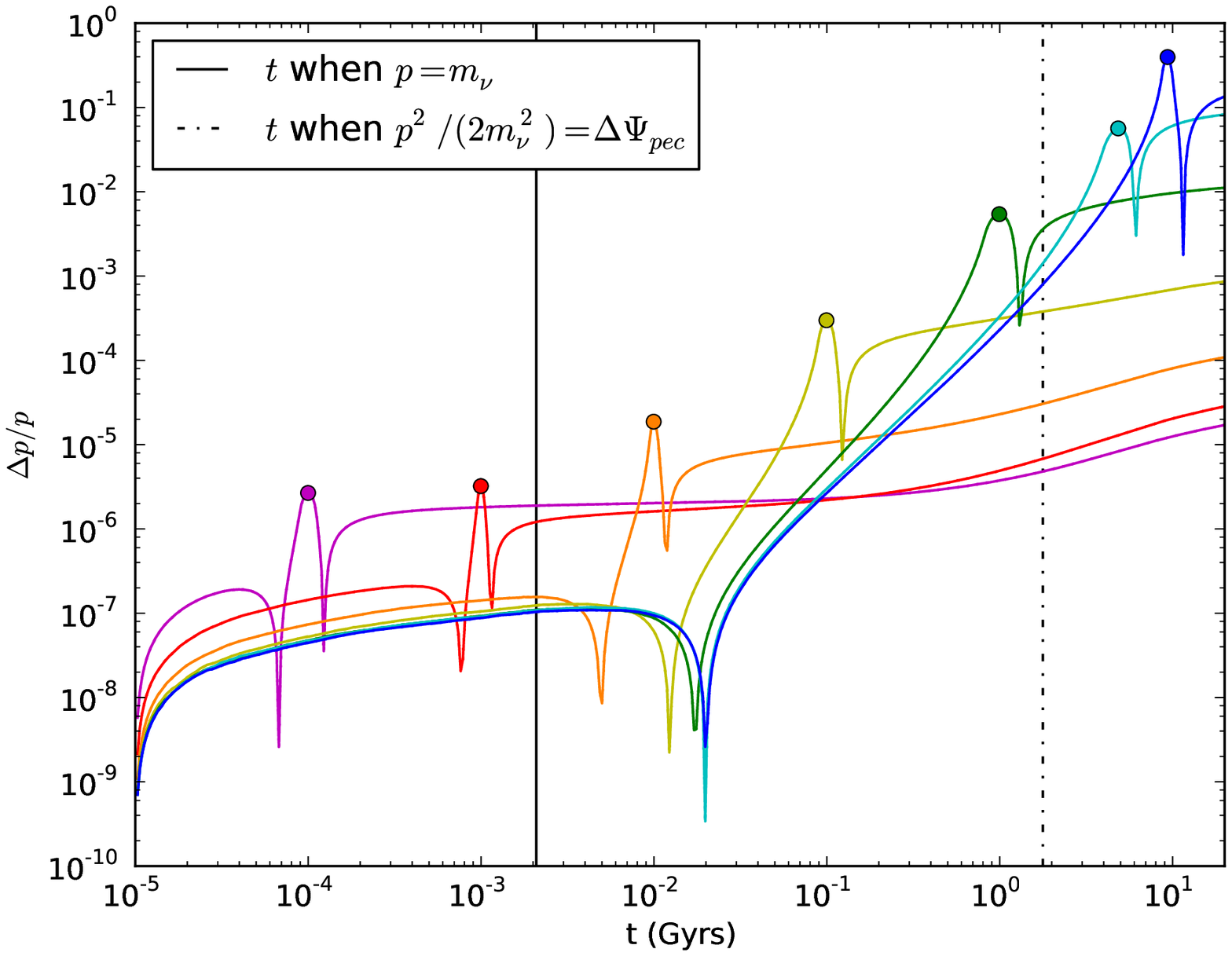} & \includegraphics[width=0.5\textwidth]{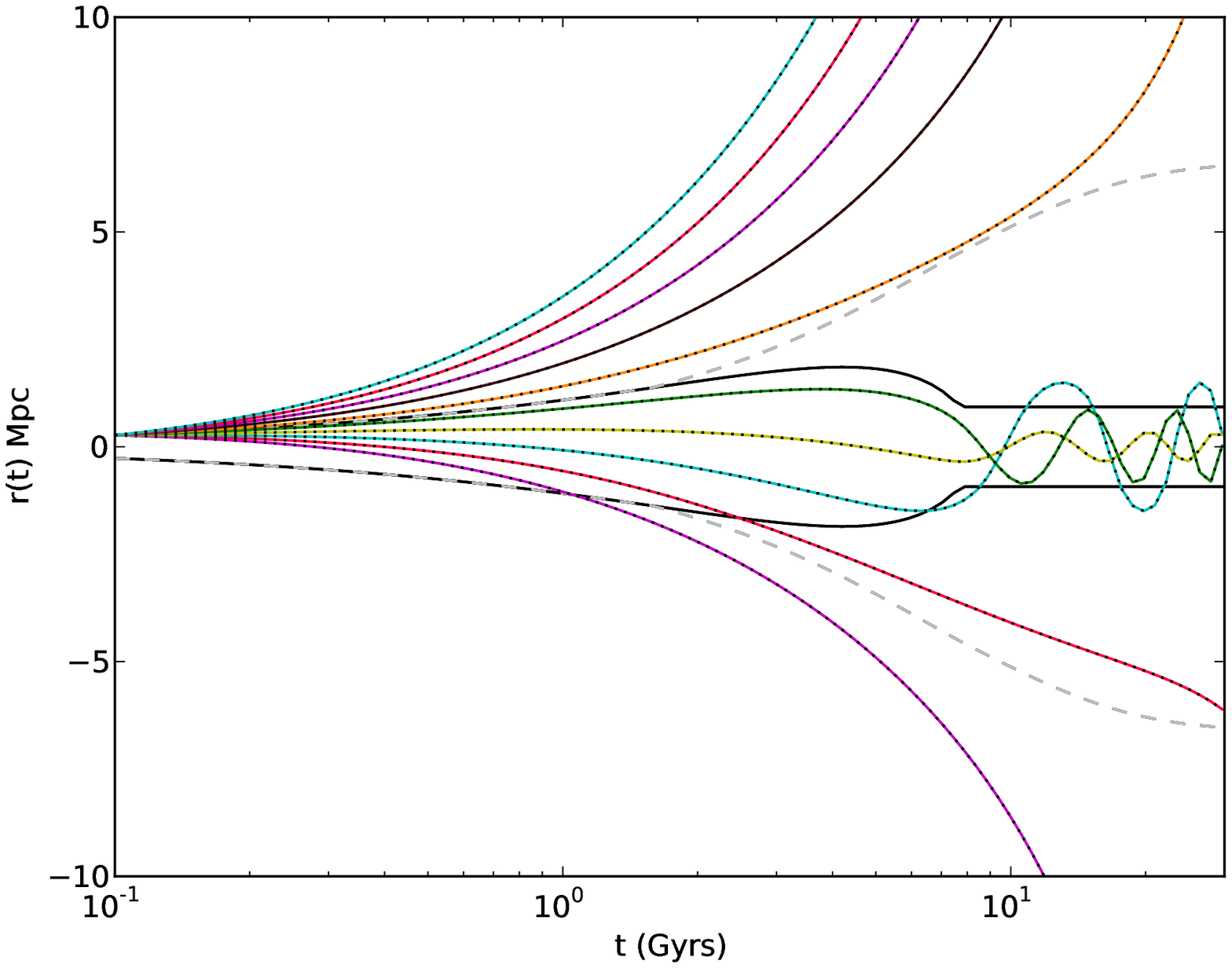} 
\end{array}$ 
\caption{\label{fig:Newtchecks} Left: Fractional change in momentum due to a potential $\Psi$ for a neutrino with $m_\nu = 0.2eV$ and $p = T_\nu/a$, calculated with Eq.~(\ref{eq:dpdt}) with the initial condition $\Delta p(t_{initial}) = 0$ (here at $t_{initial}= 10^{-5} Gyrs$). The different colors are for neutrinos encountering the potential at different times (the solid circle indicates the time when that neutrino reaches the center of the potential, $r=0$). Right: A comparison of the trajectories calculated using the full relativistic equation of motion, Eq.~(\ref{eq:RelEOM}) (shown in sold colors) and the Newtonian approximation Eq.~(\ref{eq:NREOM}) (barely visible black dotted lines appearing on top of the solid lines). The solid black line shows the radius of the halo $R_c(t)$. }
\end{center}
\end{figure}

Now we ask whether it is justified to use the Newtonian equation of motion. From Eq.~(\ref{eq:dpdt}) the change in momentum due to the gravitational potential of the halo is $\Delta p/p  \sim E^2/p^2\Delta \Psi$
where $p$, $E$ are the momentum and energy of the neutrino when it enters the potential and $\Delta \Psi = \Delta r \partial_r\Psi$. Roughly, an order unity change in the momentum (needed if the particle is to turn around) is generated when $u^2 = p^2/E^2 < \Psi$. Since $\Psi\sim 10^{-5}$, neglecting the $\mathcal{O}(u^2)$ terms is justified \footnote{A time changing potential also generates a shift in $\Delta p \sim \dot\Phi\Delta t\, p$. However, since $\dot\Phi\sim H\Phi$ this requires $\Delta t H \sim 1/\Phi \sim 10^{5}$ to cause an order unity change to $p$.}.  Figure~\ref{fig:Newtchecks} shows this explicitly: the fractional change in momentum is small for neutrinos that reach the halo with $u^2 \gsim \Psi$. In Fig.~\ref{fig:Newtchecks} we also show neutrino trajectories calculated with exact relativistic expression Eq.~(\ref{eq:RelEOM}) and the Newtonian approximation Eq.~(\ref{eq:NREOM}) -- there is no visible difference for the range of times and momenta we are interested in.

 \section{Criterion for an individual neutrino to be bound}
 \label{sec:bound}
  We want a simple condition to test whether a particle passing through $R_\nu$ or $r_*$ will remain bound to the halo (i.e. remain orbiting within $R_\nu$ or $r_*$). Define the energy per unit mass along each trajectory by
\be
\label{eq:ENewton}
E(r,v,t)=\frac{1}{2}|{\bf v}|^2-\frac{1}{2}\left(\dot{H}+H^2\right)r^2+\Psi_{pec}(r,t)\,.
\ee
 The explicit time dependence of the background, $\Psi_H(r,t)$, means that even if the peculiar gravitational potential $\Psi_{pec}$ is static, the energy of particles is not conserved.   

 A particle orbiting in the halo with orbital radius $R_\nu$ will have ${\bf v}=0$ at turnaround ($r=R_\nu$), therefore the energy of the orbit is
\ba
E_{crit} (R_\nu,t)&=& \Psi_H(R_\nu,t)+\Psi_{pec}(R_\nu,t)\\
&=&\frac{3G \delta M}{2R_c}\left(1-\frac{2}{3}\frac{R_c(t)}{R_{\nu}(t)}\left(1+\frac{\ddot{a}}{2|\ddot{a}|}\right)\right)\,.
\ea
After virialization this critical energy approaches constant values for both EdS ($\Omega_m = 1$) and $\Lambda$CDM universes: $E_{crit}\rightarrow \frac{3}{2}GM/R_c$ in EdS and $E_{crit}\rightarrow \frac{3}{2}GM/R_c(1-(8\pi \rho_\Lambda/(3 M))^{1/3}R_c)$ in $\Lambda CDM$.  A particle with orbital radius $r_*$ will have 
\ba
E_{crit}(r_*,t) = \frac{3G \delta M}{2R_c}\left(1-\frac{1}{3}\frac{R_c(t)}{r_*(t)}\left(1- \frac{3 P}{\rho}\right)\right)\,,
\ea
if $t_*$ is the time when $r(t)=r_*$ and $v_* = v(t_*)$. Note that to be entering the radius $r_*$ between $t_*$ and $t_* + \Delta t$ we need 
\be
r(t_*+\Delta t)  \le r_*(t_*+\Delta t) \rightarrow v_* \le \frac{dr_*}{dt}\,.
\ee

\begin{figure}[h!t]
\begin{center}$\begin{array}{ccc}
\includegraphics[width=0.3\textwidth]{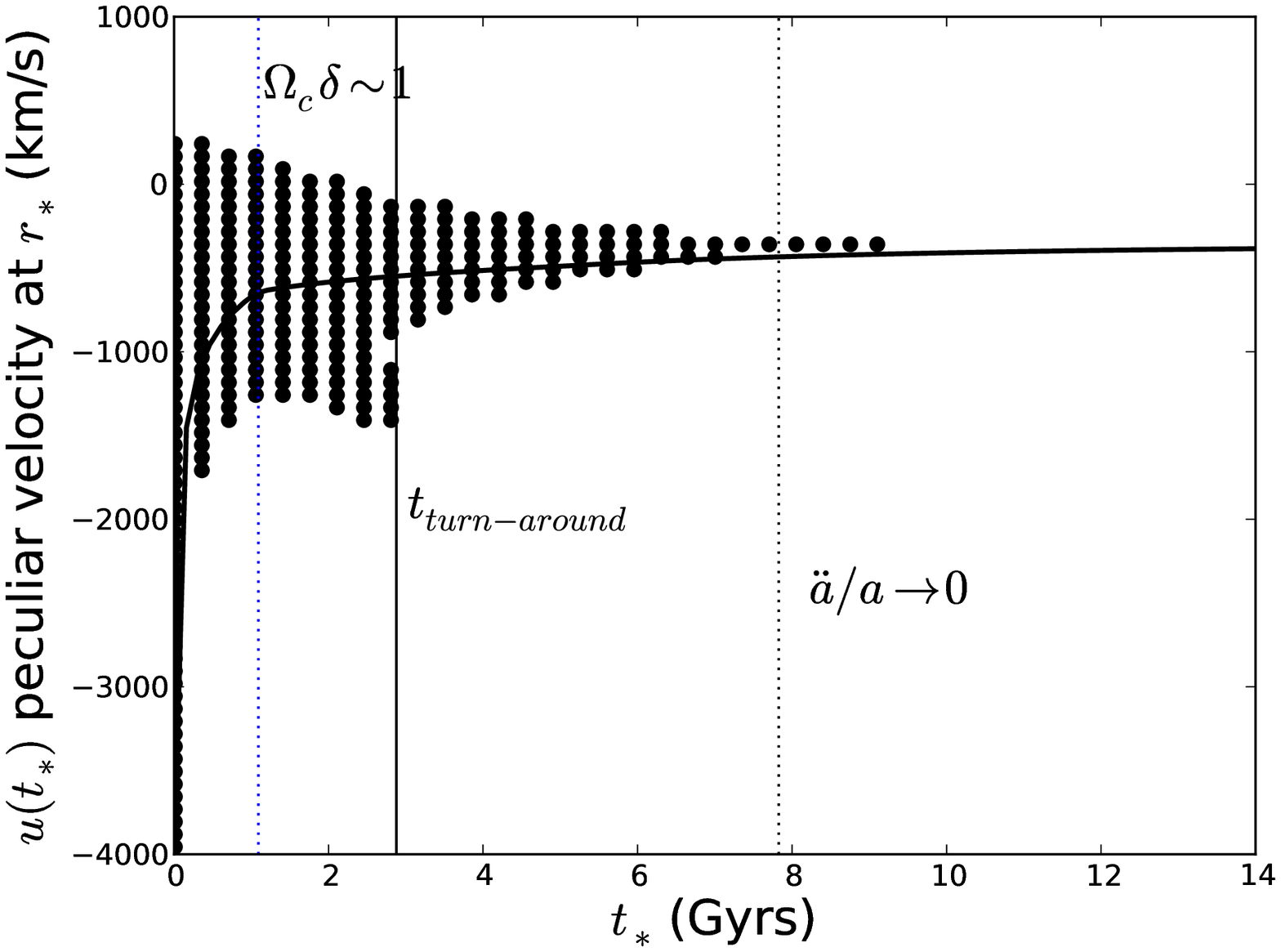} &\includegraphics[width=0.3\textwidth]{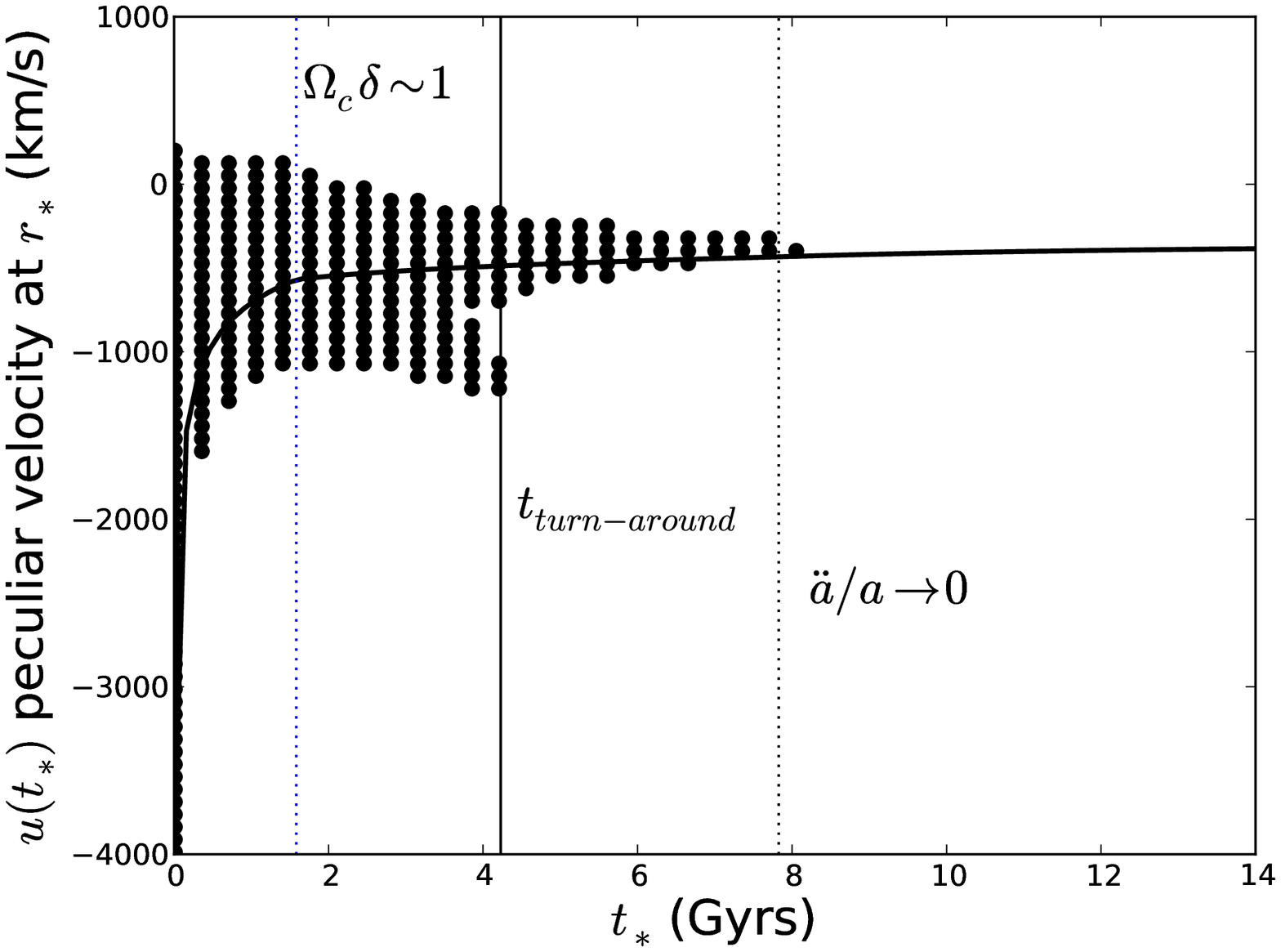} &\includegraphics[width=0.3\textwidth]{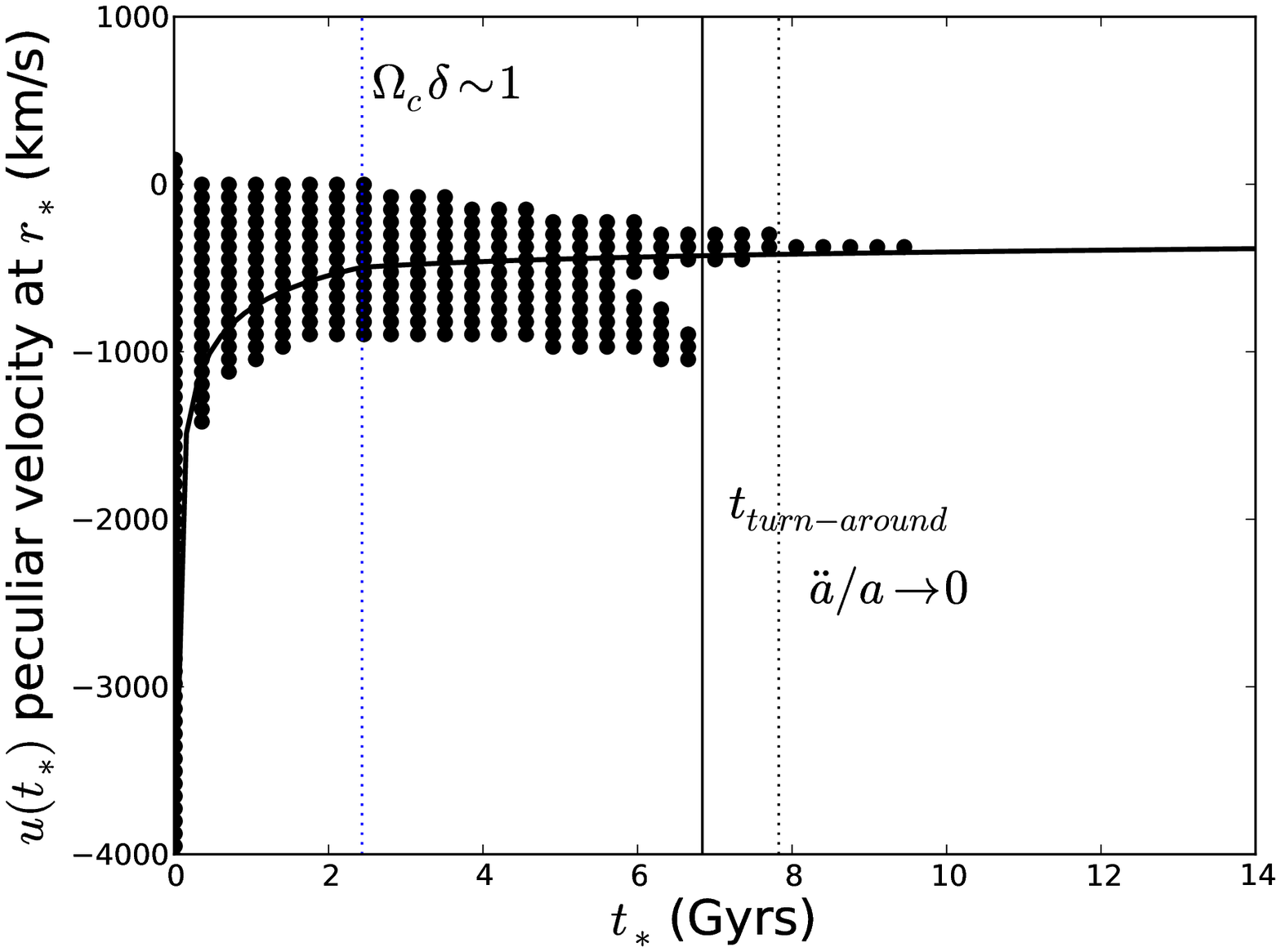} \\
\includegraphics[width=0.3\textwidth]{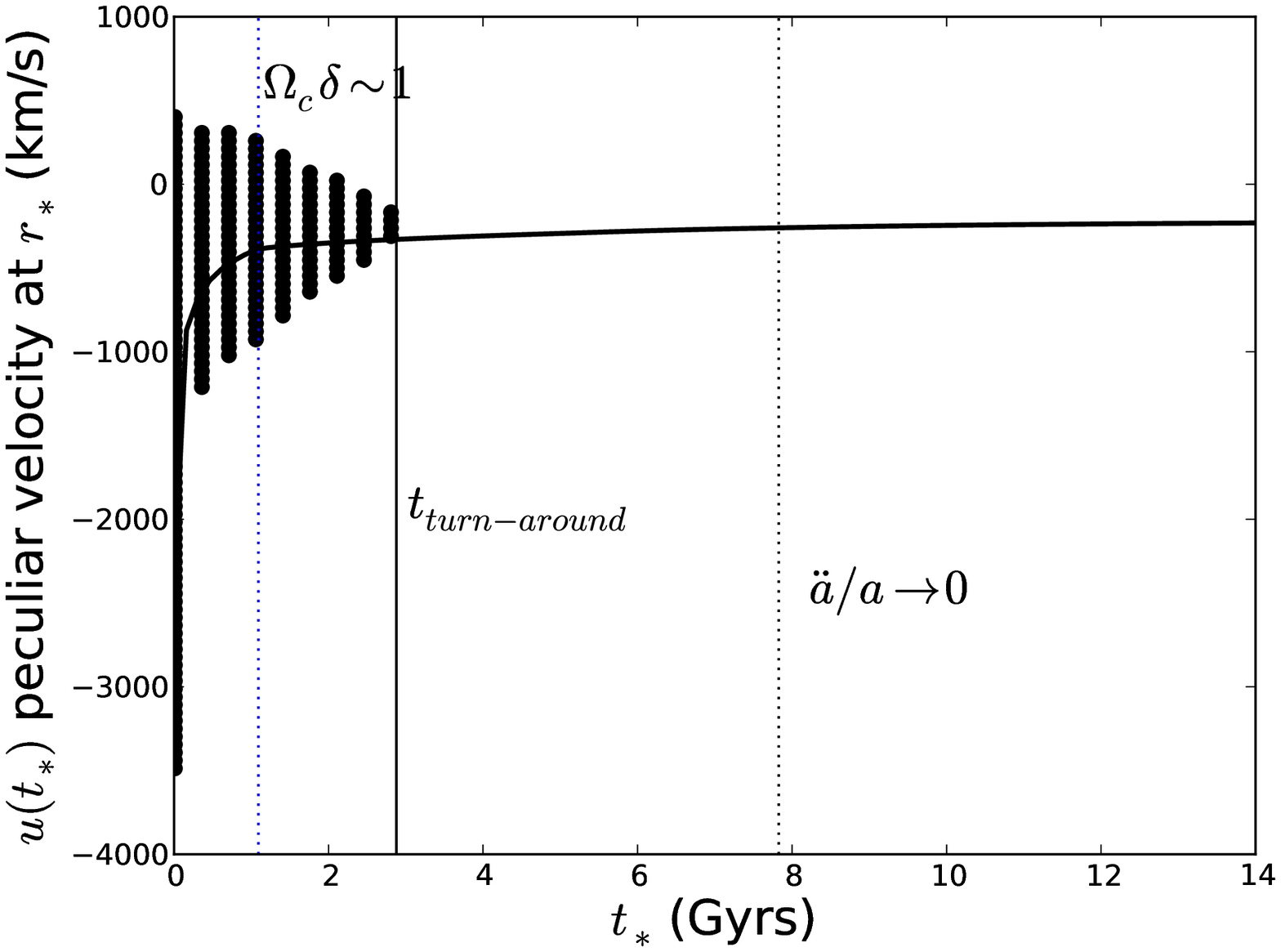} &\includegraphics[width=0.3\textwidth]{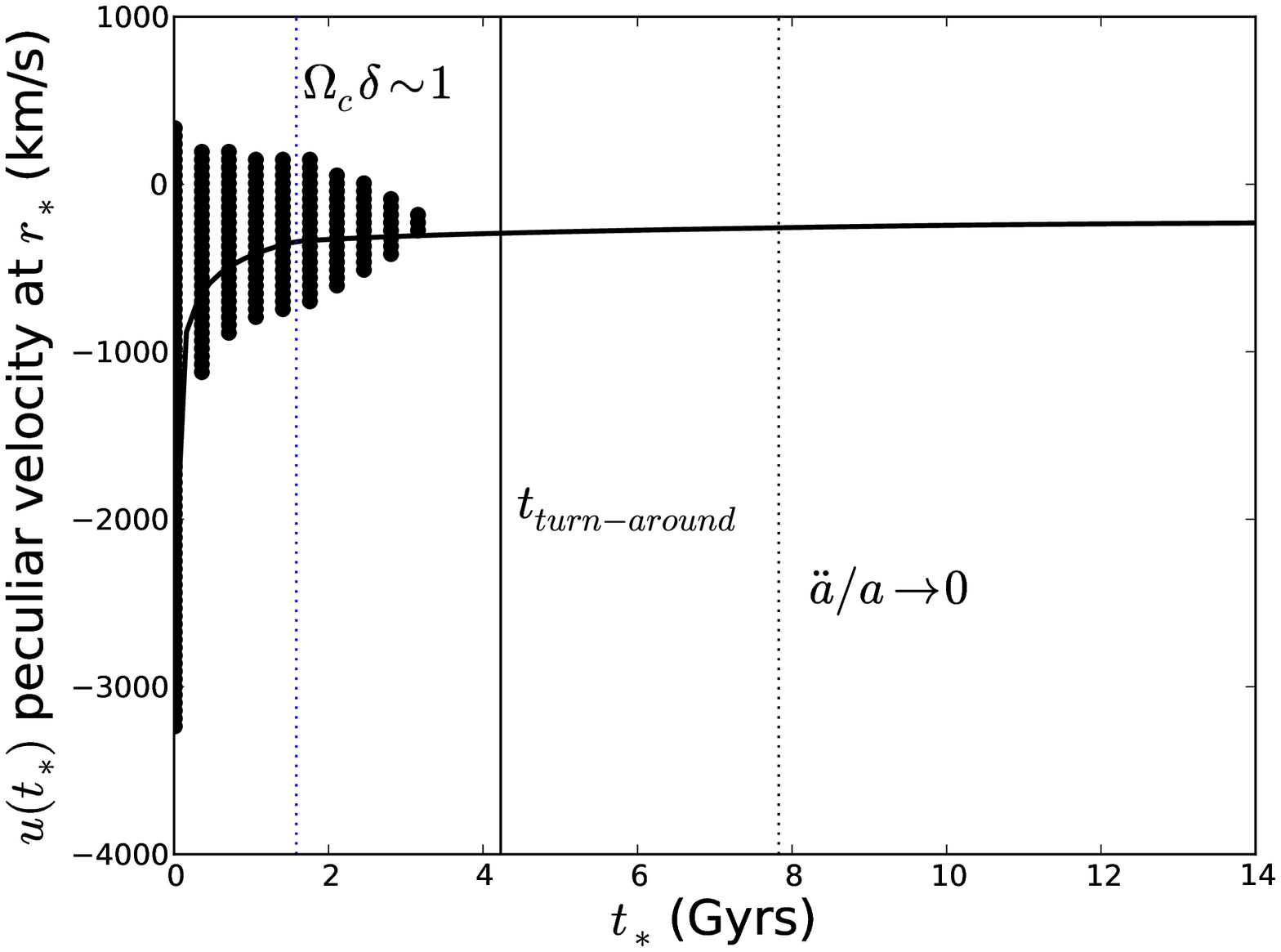} &\includegraphics[width=0.3\textwidth]{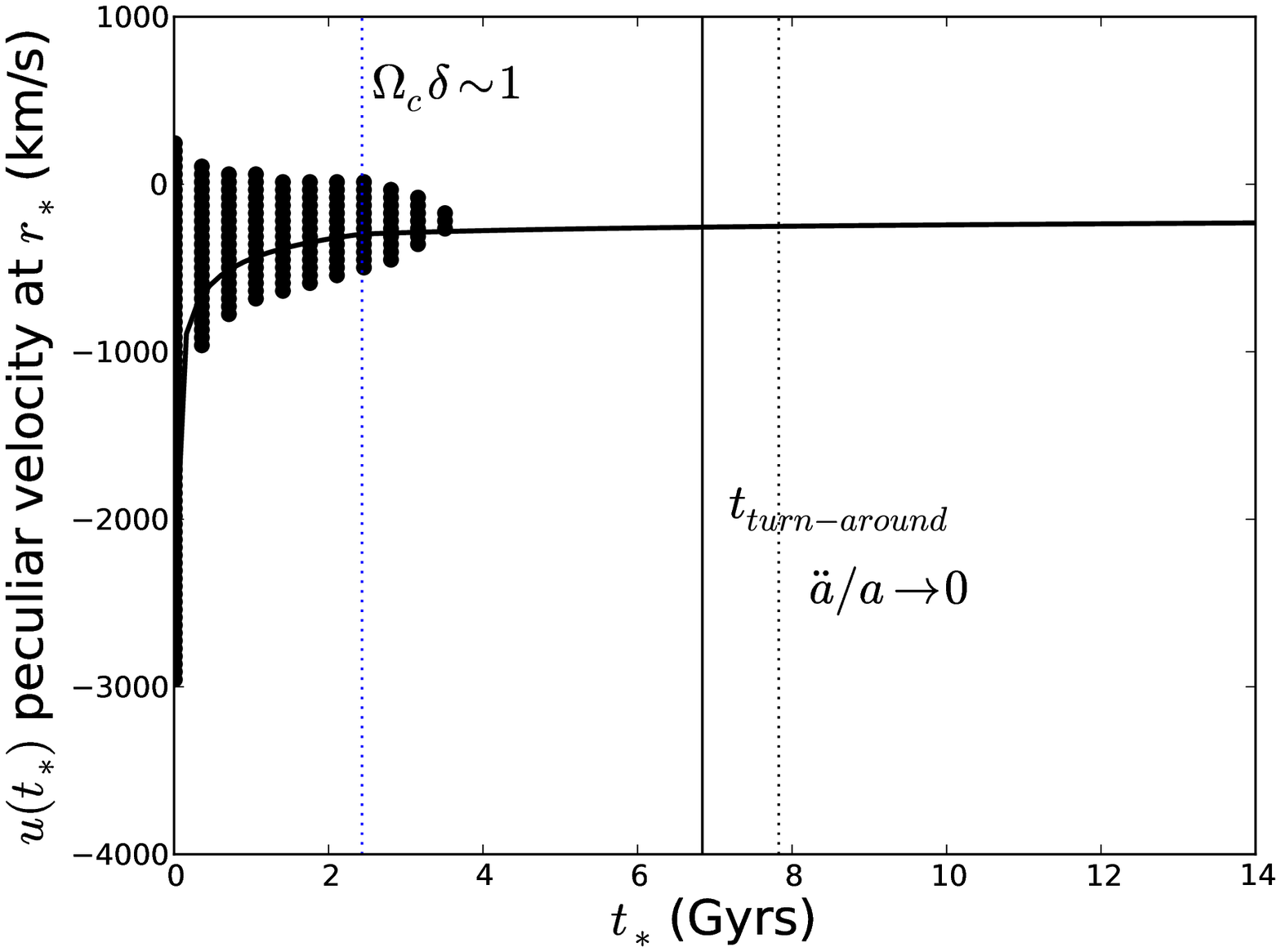} \\
\includegraphics[width=0.3\textwidth]{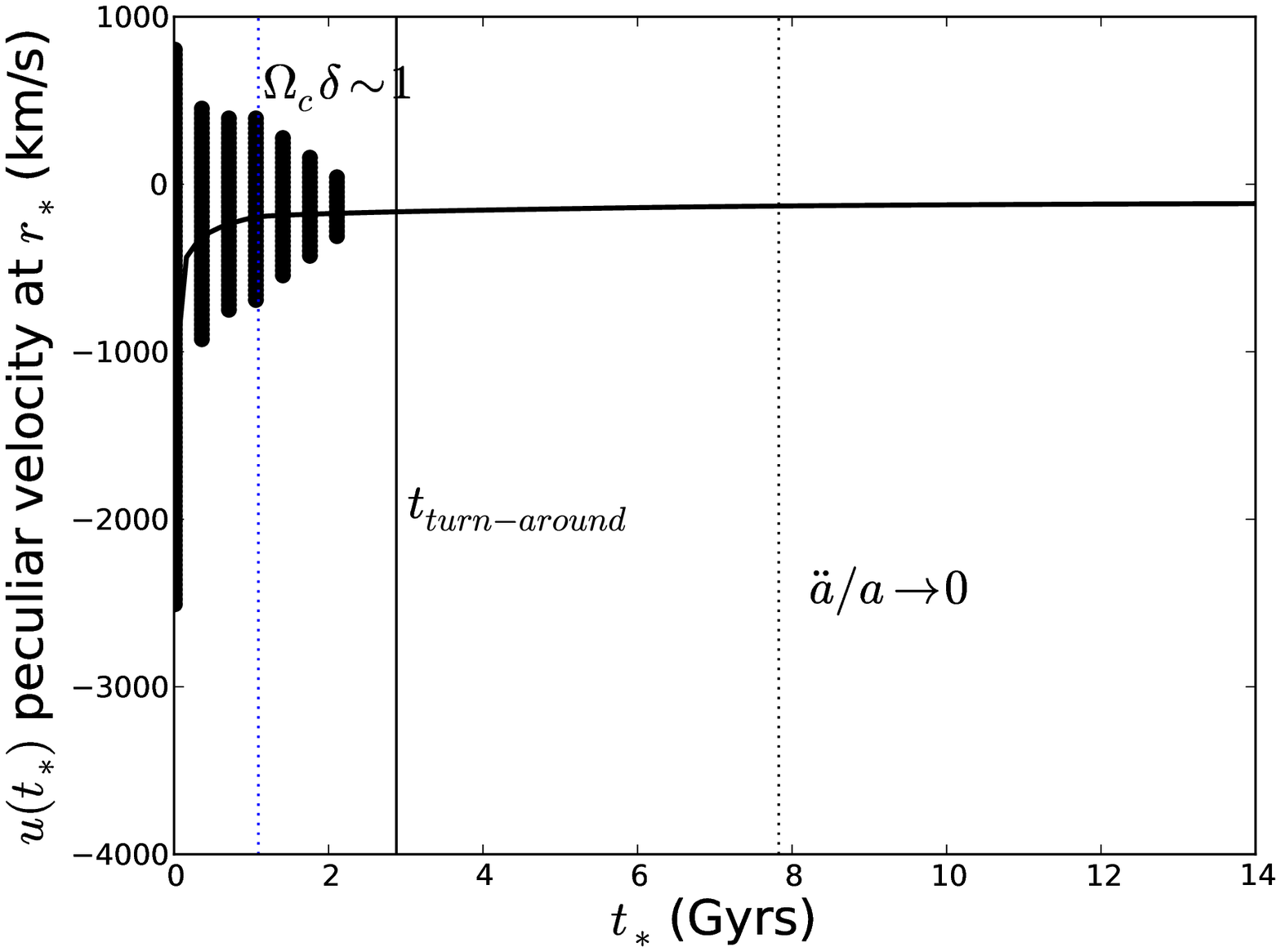} &\includegraphics[width=0.3\textwidth]{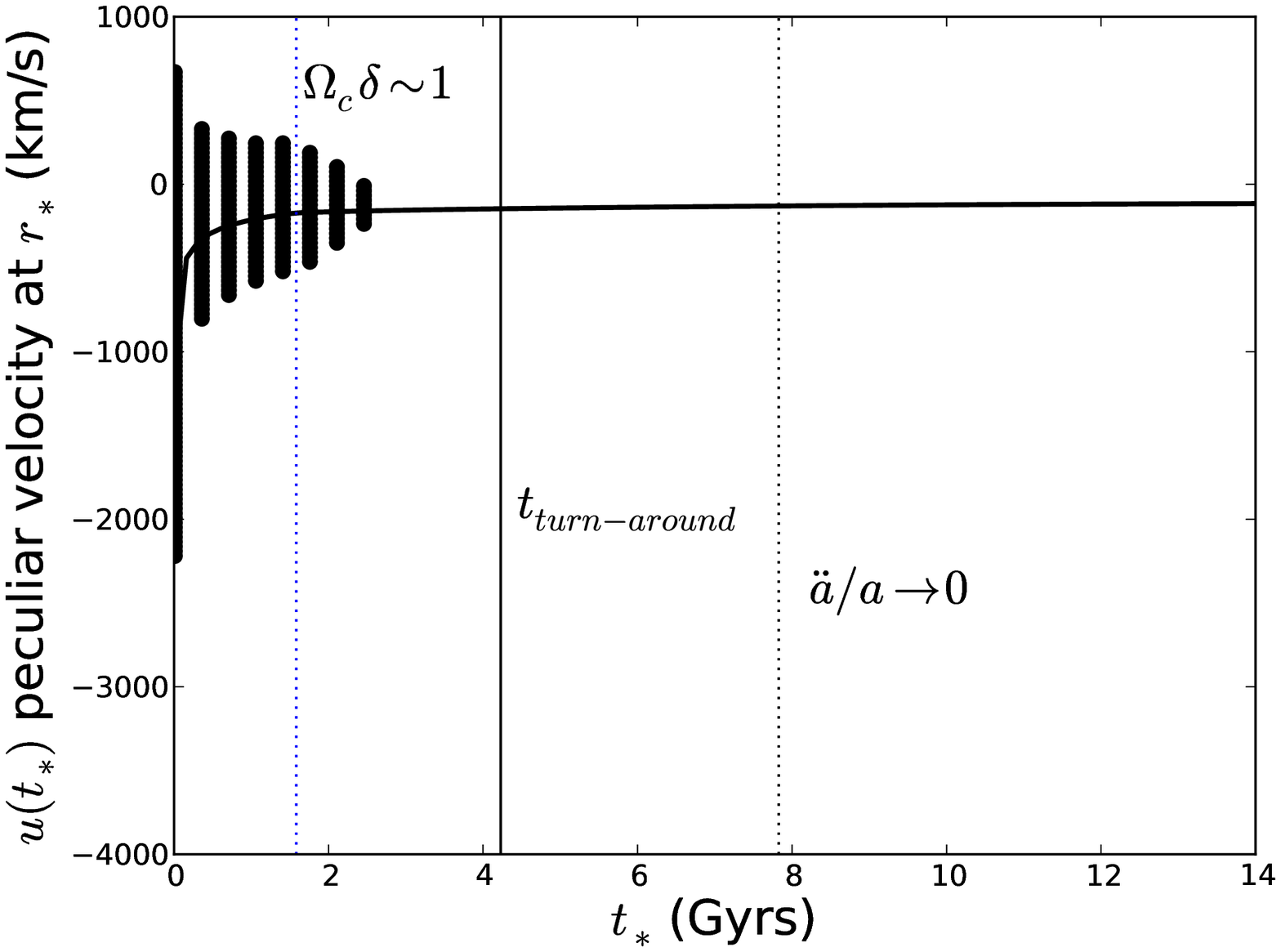} &\includegraphics[width=0.3\textwidth]{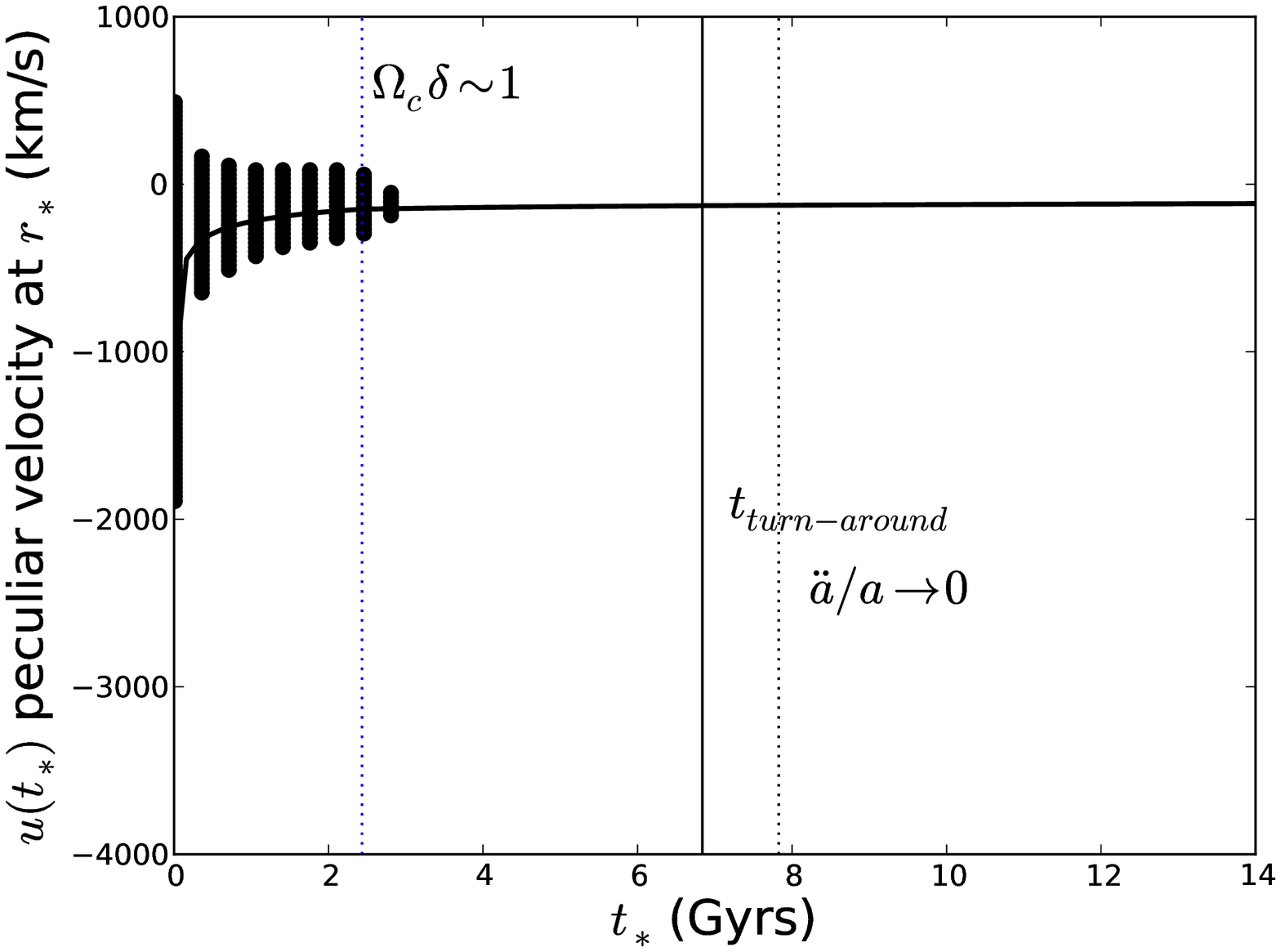} \\
\end{array}$
\caption{\label{fig:vcapture} The black dotted area indicates the range of comoving velocities of neutrinos of mass $0.2eV$ at the time they crossed $r_*$ that remain bound within the halo. Each column shows the critical velocities for a fixed halo, each halo has the same mass , $M = 10^{14} M_{\odot}$ but a different time evolution, the turn around time $\dot R_c(t_{turn-around}) \rightarrow 0$ is indicated by the solid vertical line, from left to right: $t_{turn-around}\approx 2.9\, Gyrs$, the middle $t_{turn-around} \approx 4.2 \,Gyrs$, $t_{turn-around} \approx 6.8\, Gyrs$ (corresponding to $z_{collapse} \approx 1$, $0.5$, $0$ respectively). Each row shows trajectories with fixed incident angle with respect to $\hat{{\bf r}}$. From top to bottom $\mu = 1$, $\mu = 0.6$, $\mu = 0.3$ where $\mu \equiv \hat{\bf r}\cdot \hat{\bf u}$. The black regions are centered around the line $u = r_*H(t_*)\mu$ (black curves). The time at which $\ddot{a}/a$ changes sign is shown by the vertical dotted line.  The vertical blue dotted line indicates the time at which $\Omega_c\delta \sim 1$. For $\Omega_c\delta > 1$ the peculiar velocity due to the halo is larger than the Hubble flow for $r < r_*$, but if $\Omega_c\delta <1$ the Hubble velocity dominates at all radii (see \S \ref{sec:dynamics}, Eq.~(\ref{eq:Rnu}) and Eq.~(\ref{eq:rstar})). }
\end{center}
\end{figure}

To determine the bound criterion on the peculiar velocity of particles reaching $r_*$ we numerically integrate the trajectories for neutrinos with a range of peculiar velocities and determine which neutrinos remain bound at late times. Our definition of a neutrino that is``bound at late times" is that the neutrino trajectory satisfies $r(t_{late}) < R_\nu(t_{late})$ and $E(t_{late}) < E_{crit}(R_\nu, t_{late})$. For $t_{late} \gsim 20 Gyrs$ (after which $\Omega_\Lambda \gsim 0.9$) our results are insensitive to the precise value of $t_{late}$. We then determine the range of peculiar velocities of particles crossing $r_*$ at each $t$ that are bound at late times. Our results are plotted in Fig.~\ref{fig:vcapture}. We see that, as expected,  neutrino accretion is truncated around the time when $\ddot{a}/a\rightarrow 0$. Another visible feature is that, just before $t_{turn-around}$ (the time when $R_c$ turns around in spherical collapse) the region in velocity space is disjoint. From studying individual particle trajectories, we found that the particles in the lower disjoint region are neutrinos traveling through the halo during collapse, so we attribute the appearance of the lower region to the time changing potential during collapse. 


\bibliographystyle{ieeetr}
\bibliography{mdm}

\begin{thebibliography}{10}

\bibitem{Ade:2013zuv}
P.~Ade {\em et~al.}, ``{Planck 2013 results. XVI. Cosmological parameters},''
  2013.

\bibitem{Hinshaw:2012aka}
G.~Hinshaw {\em et~al.}, ``{Nine-Year Wilkinson Microwave Anisotropy Probe
  (WMAP) Observations: Cosmological Parameter Results},'' 2012.

\bibitem{Hou:2012xq}
Z.~Hou, C.~Reichardt, K.~Story, B.~Follin, R.~Keisler, {\em et~al.},
  ``{Constraints on Cosmology from the Cosmic Microwave Background Power
  Spectrum of the 2500-square degree SPT-SZ Survey},'' 2012.

\bibitem{Sievers:2013ica}
J.~L. Sievers, R.~A. Hlozek, M.~R. Nolta, V.~Acquaviva, G.~E. Addison, {\em
  et~al.}, ``{The Atacama Cosmology Telescope: Cosmological parameters from
  three seasons of data},'' 2013.

\bibitem{Lesgourgues:2006nd}
J.~Lesgourgues and S.~Pastor, ``{Massive neutrinos and cosmology},'' {\em
  Phys.Rept.}, vol.~429, pp.~307--379, 2006.

\bibitem{Weinberg:2008zzc}
S.~Weinberg, ``{Cosmology},'' 2008.

\bibitem{Beringer:1900zz}
J.~Beringer {\em et~al.}, ``{Review of Particle Physics (RPP)},'' {\em
  Phys.Rev.}, vol.~D86, p.~010001, 2012.

\bibitem{Reid:2009nq}
B.~A. Reid, L.~Verde, R.~Jimenez, and O.~Mena, ``{Robust Neutrino Constraints
  by Combining Low Redshift Observations with the CMB},'' {\em JCAP},
  vol.~1001, p.~003, 2010.

\bibitem{Thomas:2009ae}
S.~A. Thomas, F.~B. Abdalla, and O.~Lahav, ``{Upper Bound of 0.28eV on the
  Neutrino Masses from the Largest Photometric Redshift Survey},'' {\em
  Phys.Rev.Lett.}, vol.~105, p.~031301, 2010.

\bibitem{Swanson:2010sk}
M.~E. Swanson, W.~J. Percival, and O.~Lahav, ``{Neutrino Masses from Clustering
  of Red and Blue Galaxies: A Test of Astrophysical Uncertainties},'' {\em
  Mon.Not.Roy.Astron.Soc.}, vol.~409, pp.~1100--1112, 2010.

\bibitem{Xia:2012na}
J.-Q. Xia, B.~R. Granett, M.~Viel, S.~Bird, L.~Guzzo, {\em et~al.},
  ``{Constraints on Massive Neutrinos from the CFHTLS Angular Power
  Spectrum},'' {\em JCAP}, vol.~1206, p.~010, 2012.

\bibitem{RiemerSorensen:2011fe}
S.~Riemer-Sorensen, C.~Blake, D.~Parkinson, T.~M. Davis, S.~Brough, {\em
  et~al.}, ``{The WiggleZ Dark Energy Survey: Cosmological neutrino mass
  constraint from blue high-redshift galaxies},'' {\em Phys.Rev.}, vol.~D85,
  p.~081101, 2012.

\bibitem{Zhao:2012xw}
G.-B. Zhao, S.~Saito, W.~J. Percival, A.~J. Ross, F.~Montesano, {\em et~al.},
  ``{The clustering of galaxies in the SDSS-III Baryon Oscillation
  Spectroscopic Survey: weighing the neutrino mass using the galaxy power
  spectrum of the CMASS sample},'' 2012.

\bibitem{dePutter:2012sh}
R.~de~Putter, O.~Mena, E.~Giusarma, S.~Ho, A.~Cuesta, {\em et~al.}, ``{New
  Neutrino Mass Bounds from Sloan Digital Sky Survey III Data Release 8
  Photometric Luminous Galaxies},'' {\em Astrophys.J.}, vol.~761, p.~12, 2012.

\bibitem{Wyman:2013lza}
M.~Wyman, D.~H. Rudd, R.~A. Vanderveld, and W.~Hu, ``{nu-LCDM: Neutrinos
  reconcile Planck with the Local Universe},'' 2013.

\bibitem{Bond:1980ha}
J.~Bond, G.~Efstathiou, and J.~Silk, ``{Massive Neutrinos and the Large Scale
  Structure of the Universe},'' {\em Phys.Rev.Lett.}, vol.~45, pp.~1980--1984,
  1980.

\bibitem{Davis:1981yf}
M.~Davis, M.~Lecar, C.~Pryor, and E.~Witten, ``{The Formation of Galaxies From
  Massive Neutrinos},'' {\em Astrophys.J.}, vol.~250, pp.~423--431, 1981.

\bibitem{Gunn:1972sv}
J.~E. Gunn and I.~Gott, J.~Richard, ``{On the Infall of Matter into Clusters of
  Galaxies and Some Effects on Their Evolution},'' {\em Astrophys.J.},
  vol.~176, pp.~1--19, 1972.

\bibitem{Ichiki:2011ue}
K.~Ichiki and M.~Takada, ``{The impact of massive neutrinos on the abundance of
  massive clusters},'' 2011.

\bibitem{1966ApJ...144..233G}
I.~H. {Gilbert}, ``{An Integral Equation for the Development of Irregularities
  in an Expanding Universe},'' {\em \apj}, vol.~144, p.~233, Apr. 1966.

\bibitem{Brandenberger:1987kf}
R.~H. Brandenberger, N.~Kaiser, and N.~Turok, ``{DISSIPATIONLESS CLUSTERING OF
  NEUTRINOS AROUND A COSMIC STRING LOOP},'' {\em Phys.Rev.}, vol.~D36, p.~2242,
  1987.

\bibitem{Singh:2002de}
S.~Singh and C.-P. Ma, ``{Neutrino clustering in cold dark matter halos :
  Implications for ultrahigh-energy cosmic rays},'' {\em Phys.Rev.}, vol.~D67,
  p.~023506, 2003.

\bibitem{Navarro:1995iw}
J.~F. Navarro, C.~S. Frenk, and S.~D. White, ``{The Structure of cold dark
  matter halos},'' {\em Astrophys.J.}, vol.~462, pp.~563--575, 1996.

\bibitem{Abazajian:2004zh}
K.~Abazajian, E.~R. Switzer, S.~Dodelson, K.~Heitmann, and S.~Habib, ``{The
  Nonlinear cosmological matter power spectrum with massive neutrinos. 1. The
  Halo model},'' {\em Phys.Rev.}, vol.~D71, p.~043507, 2005.

\bibitem{Saito:2009ah}
S.~Saito, M.~Takada, and A.~Taruya, ``{Nonlinear power spectrum in the presence
  of massive neutrinos: perturbation theory approach, galaxy bias and parameter
  forecasts},'' {\em Phys.Rev.}, vol.~D80, p.~083528, 2009.

\bibitem{Wong:2008ws}
Y.~Y. Wong, ``{Higher order corrections to the large scale matter power
  spectrum in the presence of massive neutrinos},'' {\em JCAP}, vol.~0810,
  p.~035, 2008.

\bibitem{Upadhye:2013ndm}
A.~Upadhye, R.~Biswas, A.~Pope, K.~Heitmann, S.~Habib, {\em et~al.},
  ``{Large-Scale Structure Formation with Massive Neutrinos and Dynamical Dark
  Energy},'' 2013.

\bibitem{Ringwald:2004np}
A.~Ringwald and Y.~Y. Wong, ``{Gravitational clustering of relic neutrinos and
  implications for their detection},'' {\em JCAP}, vol.~0412, p.~005, 2004.

\bibitem{Agarwal:2010mt}
S.~Agarwal and H.~A. Feldman, ``{The effect of massive neutrinos on the matter
  power spectrum},'' {\em Mon.Not.Roy.Astron.Soc.}, vol.~410, p.~1647, 2011.

\bibitem{Brandbyge:2008rv}
J.~Brandbyge, S.~Hannestad, T.~Haugbolle, and B.~Thomsen, ``{The Effect of
  Thermal Neutrino Motion on the Non-linear Cosmological Matter Power
  Spectrum},'' {\em JCAP}, vol.~0808, p.~020, 2008.

\bibitem{Viel:2010bn}
M.~Viel, M.~G. Haehnelt, and V.~Springel, ``{The effect of neutrinos on the
  matter distribution as probed by the Intergalactic Medium},'' {\em JCAP},
  vol.~1006, p.~015, 2010.

\bibitem{Marulli:2011he}
F.~Marulli, C.~Carbone, M.~Viel, L.~Moscardini, and A.~Cimatti, ``{Effects of
  Massive Neutrinos on the Large-Scale Structure of the Universe},'' {\em
  Mon.Not.Roy.Astron.Soc.}, vol.~418, p.~346, 2011.

\bibitem{AliHaimoud:2012vj}
Y.~Ali-Haimoud and S.~Bird, ``{An efficient implementation of massive neutrinos
  in non-linear structure formation simulations},'' 2012.

\bibitem{Bode:2000gq}
P.~Bode, J.~P. Ostriker, and N.~Turok, ``{Halo formation in warm dark matter
  models},'' {\em Astrophys.J.}, vol.~556, pp.~93--107, 2001.

\bibitem{Colin:2007bk}
P.~Colin, O.~Valenzuela, and V.~Avila-Reese, ``{On the Structure of Dark Matter
  Halos at the Damping Scale of the Power Spectrum with and without Relict
  Velocities},'' {\em Astrophys.J.}, vol.~673, pp.~203--214, 2008.

\bibitem{Brandbyge:2010ge}
J.~Brandbyge, S.~Hannestad, T.~Haugboelle, and Y.~Y. Wong, ``{Neutrinos in
  Non-linear Structure Formation - The Effect on Halo Properties},'' {\em
  JCAP}, vol.~1009, p.~014, 2010.

\bibitem{VillaescusaNavarro:2012ag}
F.~Villaescusa-Navarro, S.~Bird, C.~Pena-Garay, and M.~Viel, ``{Non-linear
  evolution of the cosmic neutrino background},'' {\em JCAP}, vol.~1303,
  p.~019, 2013.

\bibitem{Naoz:2006ye}
S.~Naoz and R.~Barkana, ``{The formation and gas content of high redshift
  galaxies and minihalos},'' {\em Mon.Not.Roy.Astron.Soc.}, vol.~377,
  pp.~667--676, 2007.

\bibitem{LoVerde:2014rxa}
M.~LoVerde, ``{Spherical collapse in $\nu \Lambda CDM$},'' 2014.

\bibitem{Bondi:1952ni}
H.~Bondi, ``{On spherically symmetrical accretion},'' {\em
  Mon.Not.Roy.Astron.Soc.}, vol.~112, p.~195, 1952.

\bibitem{Tremaine:1979we}
S.~Tremaine and J.~Gunn, ``{Dynamical Role of Light Neutral Leptons in
  Cosmology},'' {\em Phys.Rev.Lett.}, vol.~42, pp.~407--410, 1979.

\bibitem{Kull:1996nx}
A.~Kull, R.~Treumann, and H.~Bohringer, ``{Violent relaxation of
  indistinguishable objects and neutrino hot dark matter in clusters of
  galaxies},'' {\em Astrophys.J.}, vol.~466, pp.~L1--L4, 1996.

\bibitem{VillaescusaNavarro:2011ys}
F.~Villaescusa-Navarro, J.~Miralda-Escude, C.~Pena-Garay, and V.~Quilis,
  ``{Neutrino Halos in Clusters of Galaxies and their Weak Lensing
  Signature},'' {\em JCAP}, vol.~1106, p.~027, 2011.

\bibitem{Tseliakhovich:2010bj}
D.~Tseliakhovich and C.~Hirata, ``{Relative velocity of dark matter and
  baryonic fluids and the formation of the first structures},'' {\em
  Phys.Rev.}, vol.~D82, p.~083520, 2010.

\bibitem{LoVerde2014}
M.~LoVerde and M.~Zaldarriaga, ``{Effects of the relative velocity of dark
  matter and neutrinos on neutrino clustering},'' {\em in prep.}

\end{thebibliography}
\end{document}